\documentclass[prd,eqsecnum,tightenlines,preprint,nofootinbib]{revtex4-1}

\pdfoutput=1

\usepackage{graphicx}
\usepackage{amssymb}
\usepackage{amsmath}

\usepackage{slashed}

\usepackage{color}

\newcommand\eqn[1]{Eq.~\eqref{#1}}
\newcommand\sect[1]{Section~\ref{#1}}

\def\be{\begin{equation}}
\def\ee{\end{equation}}
\newcommand\ab[1]{\langle #1 \rangle}
\newcommand\sqb[1]{[ #1 ]}

\def\mA{\mathcal A}
\def\mN{\mathcal N}
\def\tf{\tilde f}
\def\tr{\textrm{tr}}
\newcommand\trf[1]{\text{tr}_5 (#1)}

\def\nn{\nonumber}

\begin{document}

\title{Integrand Oxidation and One-Loop Colour-Dual Numerators \\ in $\mN=4$ Gauge Theory}

\author{N. Emil J. Bjerrum-Bohr}
\author{Tristan Dennen}
\author{Ricardo Monteiro}
\author{Donal O'Connell}
\affiliation{Niels Bohr International Academy and Discovery Center, \\ Niels Bohr Institute, University of Copenhagen, Blegdamsvej 17, DK-2100, Copenhagen \O, Denmark}

\date{\today}

\begin{abstract}
We present a systematic method to determine BCJ numerators for one-loop amplitudes that explores the global constraints on the loop momentum dependence. We apply this method to amplitudes in $\mN=4$ gauge theory, working out detailed examples up to seven points in both the MHV and the NMHV sectors. We see no obstruction to the application of our method to higher point one-loop amplitudes. The structure of Jacobi identities between BCJ numerators is seen to be closely connected to that of algebraic integrand reductions. We discuss the consequences for one-loop $\mN=8$ supergravity amplitudes obtained through the double copy prescription. Moreover, in the MHV sector, we show how to obtain simple BCJ box numerators using a relationship with amplitudes in self-dual gauge theory. We also introduce simpler trace-type formulas for integrand reductions.

\end{abstract}

\maketitle
\tableofcontents



\section{Introduction}
\label{IntroductionSection}

In 2008, Bern, Carrasco, and Johansson uncovered a curious property of tree-level amplitudes in $\mN=4$ supersymmetric Yang-Mills (SYM) theory~\cite{Bern:2008qj}. They noticed that it is possible to rearrange the amplitudes into a form where kinematic factors and colour factors appear symmetrically, and where the kinematic factors satisfy the same algebraic identities as the colour factors. This property has come to be known as the BCJ duality between colour and kinematics, and it suggests the presence of a hidden symmetry in a wide range of Yang-Mills theories. Moreover, one can construct certain gravity amplitudes by replacing colour factors with another copy of BCJ dual kinematic factors, exposing a mysterious connection between the structure of amplitudes in Yang-Mills and gravity theories. 

Since the BCJ paper appeared, many interesting features of the tree-level duality have been found~\cite{Sondergaard:2009za,Tye:2010dd, BjerrumBohr:2010zs, Bern:2010yg,Vaman:2010ez, Bern:2011ia, Mafra:2011kj, Monteiro:2011pc,Du:2011js,Broedel:2011pd, BjerrumBohr:2012mg, Fu:2012uy}. The so-called BCJ amplitude relations, new linear relations among tree-level partial amplitudes that follow from the duality, add to the previously known Kleiss-Kuijf relations~\cite{Kleiss:1988ne} and have been proven using a variety of methods \cite{BjerrumBohr:2009rd,Stieberger:2009hq,Feng:2010my,Jia:2010nz,Cachazo:2012uq}. The double copy relation to gravity has also been proven at tree level~\cite{Bern:2010yg}; it is equivalent to the celebrated KLT relation~\cite{Kawai:1985xq,Bern:1998sv}, proven in field theory for the first time in~\cite{BjerrumBohr:2010ta}; see also~\cite{BjerrumBohr:2010hn,Sondergaard:2011iv}.

However, what is most striking about the BCJ duality in Yang-Mills theory and its double copy relation to gravity is that they seem to hold also at loop level~\cite{Bern:2010ue}. They have proven to be a very useful tool for studying multi-loop amplitudes in $\mN=8$ supergravity~\cite{Bern:2010ue,Bern:2012uf,Carrasco:2011mn,Vanhove:2010nf}, $\mN=4$ supergravity~\cite{Bern:2011rj,BoucherVeronneau:2011qv,Bern:2012cd,Bern:2012gh}, and also theories with less supersymmetry~\cite{Carrasco:2012ca}. Additional evidence for this loop-level structure appears in soft~\cite{Oxburgh:2012zr} and high-energy~\cite{Saotome:2012vy} limits, and also in the BCFW shifts of gauge theory integrands~\cite{Boels:2012sy}. Relations between gauge theory and gravity at one loop were also investigated in \cite{Naculich:2011fw}, and a one-loop version of the BCJ amplitude relations has been proven in~\cite{Boels:2011tp}.

Because of the connection between kinematics and colour inherent in the BCJ duality, one might expect an underlying kinematic symmetry to be at play. Indeed, a partial understanding of the kinematic symmetry has appeared in refs.~\cite{Monteiro:2011pc, BjerrumBohr:2012mg}, and this was later used to construct a BCJ representation for finite one-loop amplitudes in pure Yang-Mills theory~\cite{Boels:2013bi}. Beyond Yang-Mills scattering amplitudes, the duality has also appeared in multi-loop form factors~\cite{Boels:2012ew} and higher-dimension operators~\cite{Broedel:2012rc}. There is even evidence of the duality holding in ABJM theory, where the analogue of colour is associated to a 3-algebra instead of a (Lie) 2-algebra~\cite{Bargheer:2012gv, Huang:2012wr}.

One challenge to further progress in the gauge theory amplitude context is that it is very difficult to obtain BCJ dual forms of amplitudes at high loop orders and high multiplicities. There is no known construction in general; instead, the usual strategy is to start with an ansatz for a set of `master' numerators, from which all other numerators can be obtained through Jacobi identities. One then fixes unknown coefficients by matching the ansatz to unitarity cuts of the amplitude~\cite{Bern:1994zx, Bern:1994cg, Bern:1996ja, Bern:1995db}. The ansatz grows combinatorially as the loop order and multiplicity are increased, which puts a practical limit on the problems this can be applied to. Moreover, it is always possible that the ansatz is not general enough to contain the actual BCJ form of the amplitude.

For these reasons, we develop here a new strategy for obtaining BCJ numerators constructively rather than by ansatz. Our strategy, which we apply to $\mN=4$ SYM amplitudes, utilizes global properties of the system of one-loop numerator Jacobi identities to invert the usual top-down approach; instead of starting with master numerators and constructing all other numerators via Jacobi identities, we start with the much simpler box numerators and construct all other numerators in a bottom-up manner, a procedure we term ``integrand oxidation.'' This still assumes knowledge of the BCJ box numerators, and we present a method for constructing these in the case of MHV amplitudes. Moreover, in the absence of such BCJ box numerators, for example for non-MHV amplitudes, we present a method starting from unitarity cuts of the amplitude.

This paper is organised as follows: In~\sect{BCJReviewSection}, we review the BCJ duality between colour and kinematics. In~\sect{FivePointEasySection}, we tackle the case of the five-point one-loop MHV amplitude in $\mN=4$ SYM theory. Although this has been thoroughly discussed elsewhere~\cite{Carrasco:2011mn,Yuan:2012rg,Cachazo:2008vp}, we take the opportunity to introduce our general strategy with the simplest possible example. We see explicitly that knowledge of BCJ boxes in this case is enough to fix the entire amplitude. In Sections~\ref{SelfDualSection}~and~\ref{ReductionsSection}, we discuss two ingredients that will be necessary to progress to higher-point amplitudes. First, we give an explicit construction of BCJ box numerators by exploiting a dimension-shifting relationship between one-loop MHV amplitudes in $\mN=4$ SYM theory and one-loop all-plus amplitudes in pure Yang-Mills~\cite{Bern:1996ja}. Then we discuss scalar integral reduction techniques, presenting simplified expressions. In~\sect{SixPointSection}, we use the six-point box numerators obtained in~\sect{SelfDualSection} to reconstruct the pentagon and hexagon BCJ numerators, including all loop momentum dependence. We build on this result in~\sect{SevenPointSection} by constructing the BCJ form of the seven-point MHV amplitude in a similar manner before progressing to non-MHV six- and seven-point amplitudes in \sect{NMHVSection}. We wrap up with a discussion of how cancellations in gravity amplitudes arise from the BCJ form of the amplitude in \sect{Gravity}, and present our conclusions in \sect{ConclusionSection}.

\section{Review}
\label{BCJReviewSection}

The statement of the BCJ colour-kinematics duality is that gauge theory amplitudes can always be written as a sum over cubic Feynman-like diagrams, whose kinematic dependence mirrors the algebraic properties of the colour factors. Consider tree-level amplitudes. Associated with each cubic diagram $\Gamma_i$ are three objects: a colour factor $c_i$, a kinematic numerator $n_i$, and a set of propagator denominators $P_{ki}$ where $k$ runs over the set of propagators in $\Gamma_i$. The amplitude is 
\be 
\label{eq:ReviewTreeAmplitude}
\displaystyle{\cal A}_m\;=\; g^{m-2} \sum_{\mathrm{diagrams} \;\; \Gamma_i} \frac{n_i c_i}{\prod_{k} P_{ki}}\,. \ee
Comparing with the usual Feynman diagram expansion in gauge theory, one can see that to arrive at \eqn{eq:ReviewTreeAmplitude}, one must absorb diagrams involving four-point vertices into the numerators $n_i$. This can always be achieved by multiplying and dividing contributions by a propagator factor. This is not a unique procedure, so there are many sets of numerators such that \eqn{eq:ReviewTreeAmplitude} is a valid expression for the amplitude.
The key discovery of BCJ was that certain representations originating from such reshufflings are special. They involve numerators $n_i$ satisfying
\be \displaystyle n_i \pm n_j \pm n_k =0\,,\ee
whenever the associated colour factors satisfy a Jacobi identity
\be \displaystyle c_i \pm c_j \pm c_k =0\,.\ee
A set of numerators $n_i$ which satisfy all the Jacobi relations is said to be colour-kinematics dual; for brevity, we will also call them simply dual numerators.

It is a remarkable fact~\cite{Bern:2008qj,Bern:2010yg} that replacing the colour factors $c_i$ in \eqn{eq:ReviewTreeAmplitude} with a set $\tilde n_i$ of kinematic numerators leads to something useful as well, namely a representation of gravity directly factorized as a double copy of gauge theories:
\be {\cal M}_m\;=\; i (\kappa/2)^{m-2} \sum_{\mathrm{diagrams} \;\; \Gamma_i} \frac{n_i \tilde n_i}{\prod_{k} P_{ki}}\,. \ee
The second set of numerators, $\tilde n_i$ in this double-copy equation need not correspond to the same gauge theory. For example, one can choose $n_i$ to be a set of numerators appropriate for $\mathcal{N} = 4$ super-Yang-Mills, while $\tilde n_i$ are numerators for non-supersymmetric gauge theory. The resulting gravitational theory in this example is $\mathcal{N}=4$ supergravity. The relationships among pairs of gauge theories and a gravity theory established by the double-copy have recently been explored~\cite{Damgaard:2012fb, Carrasco:2012ca}.

The story we have discussed so far was at tree level. The colour-kinematics duality and the double copy are conjectured to extend to all loops~\cite{Bern:2010ue}, but beyond tree level less is known. At $L$ loop level, we still express $m$-point amplitudes $\mathcal{A}_m^L$ as a sum over cubic diagrams,
\be 
\label{eq:ReviewLoopAmplitude}
{\cal A}^L_m\;=\;i^{L} g^{m -2 + 2L}\sum_{\mathrm{diagrams}  \;\; \Gamma_i} \int \prod_{j=1}^L \frac{d^D \ell_j}{(2\pi)^D} \frac{1}{S_i}\frac{n_i(\ell_j) c_i}{\prod_{k} P_{ki}(\ell_j)}\,,
\ee
where $S_i$ is a symmetry factor. Just as at tree level, the condition for colour-kinematics duality is 
\begin{equation}
c_i \pm c_j \pm c_k = 0 \Rightarrow n_i(\ell) \pm n_j(\ell) \pm n_k(\ell) =0\,.
\end{equation}
The loop momenta in each numerator are chosen so that the four legs involved in the Jacobi identity have the same momentum in the three diagrams. Gravity amplitudes are then expressed as
\be 
\label{eq:loopdoublecopy}
{\cal M}^L_m\;=\;i^{L+1} (\kappa/2)^{m -2 + 2L}\sum_{\mathrm{diagrams}  \;\; \Gamma_i} \int \prod_{j=1}^L \frac{d^D \ell_j}{(2\pi)^D} \frac{1}{S_i}\frac{n_i(\ell_j) \tilde n_i(\ell_j)}{\prod_{k} P_{ki}(\ell_j)}\,.
\ee

There is, as yet, no proof that dual numerators always exist at loop level for any theory. However, numerators have been discovered in examples. In the case of the four-particle amplitude in $\mathcal{N}=4$ gauge theory, dual numerators exist for up to four loops~\cite{Bern:2010ue,Bern:2012uf}. The five point amplitude also exists at up to two loops in the same theory~\cite{Carrasco:2011mn}. In the context of studying the interplay between the double-copy construction and orbifolds in supergravity, the one-loop four point amplitudes of both $\mathcal{N} = 1$ and $\mathcal{N} = 2$ gauge theory were recently calculated~\cite{Carrasco:2012ca}. The first examples of infinite families of one-loop amplitudes in colour-kinematics dual form were discussed in~\cite{Boels:2013bi}. The relevant families of amplitudes are the helicity-equal amplitudes and amplitudes which have $m-1$ particles of one helicity, and $1$ particle of the other, in pure gauge theory. It is also known that, at two loops, the four particle helicity-equal amplitude can be put in a dual form~\cite{Bern:2010ue}.

Our focus in this paper will be on constructing colour-dual numerators for one-loop amplitudes in maximally supersymmetric gauge theory and gravity. We now turn to this topic in the simplest non-trivial case, namely the five-point amplitude.


\section{An Invitation: The Five-Point Amplitude}
\label{FivePointEasySection}

A set of dual numerators at five points was first found by Carrasco and Johansson~\cite{Carrasco:2011mn} and was more recently discussed by Yuan~\cite{Yuan:2012rg}. We will re-derive this set of numerators in a manner that will extend to higher-point amplitudes.

Suppose we know the BCJ box numerators for this amplitude; for example, we can compute box cuts in self-dual Yang-Mills theory and extract candidate MHV box numerators, as will be described in \sect{SelfDualSection}. We will label the numerator of the pentagon with legs ordering $(a,b,c,d,e)$ as $n_5(a,b,c,d,e)$; then the BCJ box numerators are given by\footnote{The subscript on $n_m$ indicates that we are talking about the BCJ numerator of an $m$-gon, the precise identity of which can be read off from the pattern of square brackets surrounding the arguments of $n_m(\ldots)$. Although the subscript is redundant notation, we leave it for clarity; however, we stress that the use of $n_5$, for example, may indicate different quantities depending on the particular amplitude under discussion.}
\begin{align}
\label{eq:5ptSwap1}
n_4([1,2],3,4,5) &= n_5(1,2,3,4,5) - n_5(2,1,3,4,5) \\
&= i \delta^8(Q) \frac{\sqb{12}^2 \sqb{34}\sqb{45}\sqb{53}}{\trf{1234}}\,,
\label{eq:5ptboxnum}
\end{align}
where $\trf{1234}$ is the usual trace over gamma matrices, $\trf{1234} \equiv \tr(\gamma_5 \slashed{p}_1 \slashed{p}_2 \slashed{p}_3 \slashed{p}_4)$, and $\delta^8(Q)$ is the supersymmetric delta function, which gives rise to the gluon MHV factor $\langle ij\rangle^4$, $i$ and $j$ being the negative helicity gluons.
This set of box numerators has the important properties that it does not depend on loop momentum, and that it is symmetric for permutations of its corners, which implies the vanishing of triangle and bubble numerators.

The Carrasco-Johansson set of numerators is independent of loop momentum. We do not expect the same simplification to generalise to higher points (and we will confirm this expectation below). Let us therefore open with a set of numerators at five points which may depend linearly on the loop momentum $\ell$:
\begin{equation}
n_5(1,2,3,4,5;\ell) = n_{1,5}(1,2,3,4,5)\cdot\ell + n_{0,5}(1,2,3,4,5),
\end{equation}
where we choose our loop momentum to be between the last and first argument of the numerator. That is,
\begin{equation}
n_5(1,2,3,4,5;\ell) \leftrightarrow 
\begin{minipage}[c]{0.2\linewidth}
\centering
\includegraphics[scale=0.3]{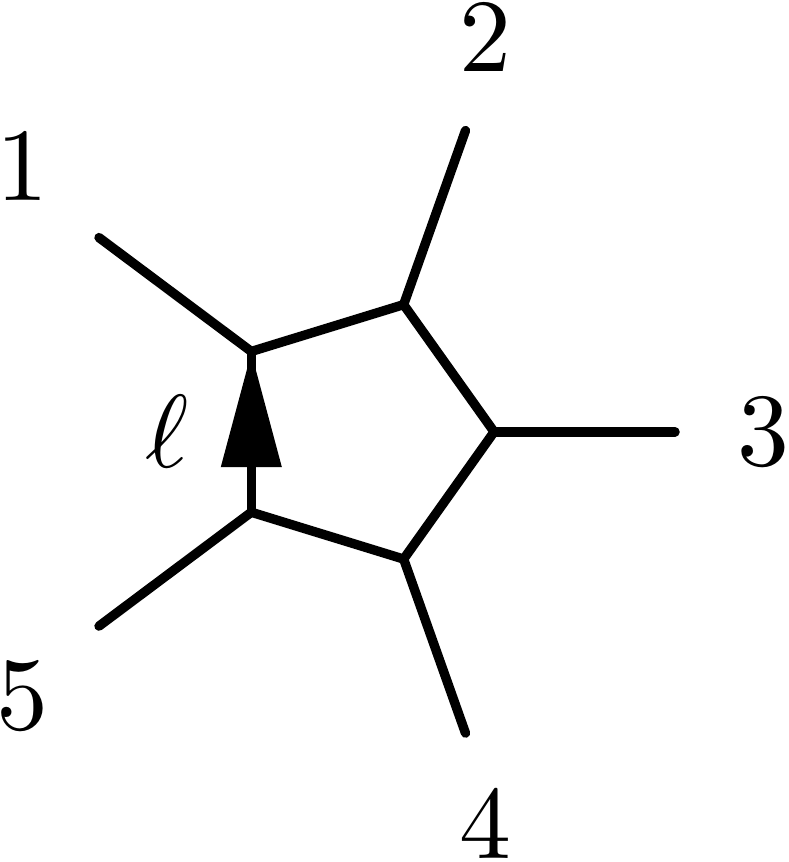}
\end{minipage}.
\end{equation}
The basis of our method is to explore the constraints on the loop momentum dependence coming from the Jacobi identities. Consider the following set of equations, obtained by using \eqn{eq:5ptSwap1} iteratively to move leg $1$ around the loop:
\begin{align}
n_4(2,[1,3],4,5) &= n_5(2,1,3,4,5;\ell) - n_5(2,3,1,4,5;\ell) \,,\nn\\
n_4(2,3,[1,4],5) &= n_5(2,3,1,4,5;\ell) - n_5(2,3,4,1,5;\ell) \,,\nn\\
n_4(2,3,4,[1,5]) &= n_5(2,3,4,1,5;\ell) - n_5(2,3,4,5,1;\ell) \,.
\end{align}
Summing these equations, we learn that
\begin{align}
\label{eq:5ptwalking}
 n_4([1,2],3,4,5) &+ n_4(2,[1,3],4,5) + n_4(2,3,[1,4],5) + n_4(2,3,4,[1,5]) \nn\\
& = n_5(1,2,3,4,5;\ell) - n_5(2,3,4,5,1;\ell) \,.
\end{align}
However, the pentagon numerator must satisfy the cyclic permutation property,
\begin{equation}
n_5(2,3,4,5,1;\ell) \leftrightarrow
\begin{minipage}[c]{0.2\linewidth}
\centering
\includegraphics[scale=0.3]{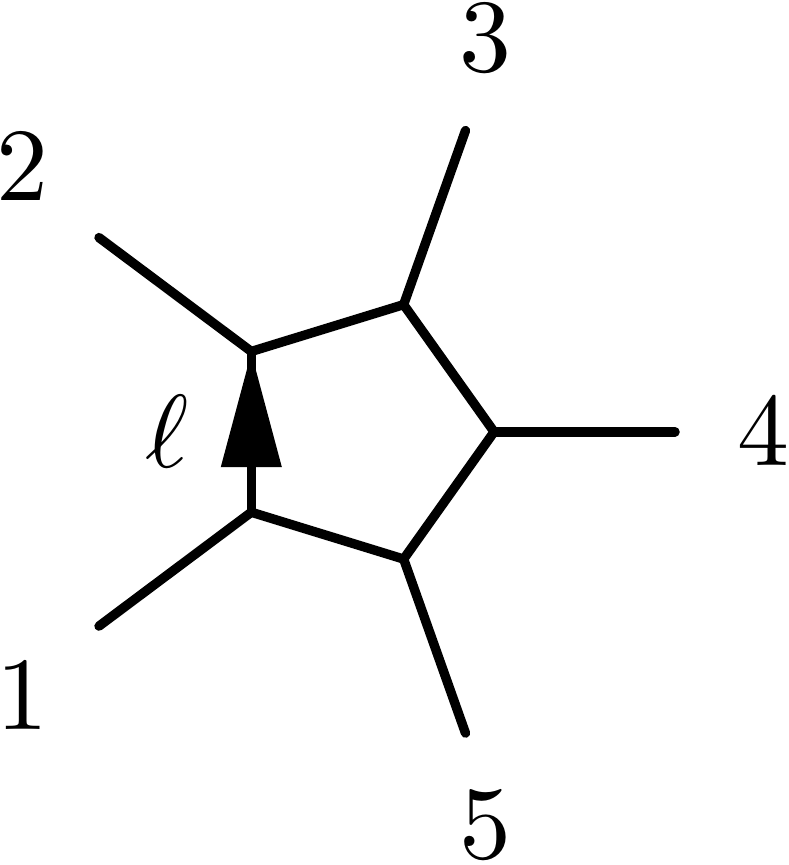}
\end{minipage} = 
\begin{minipage}[c]{0.2\linewidth}
\centering
\includegraphics[scale=0.3]{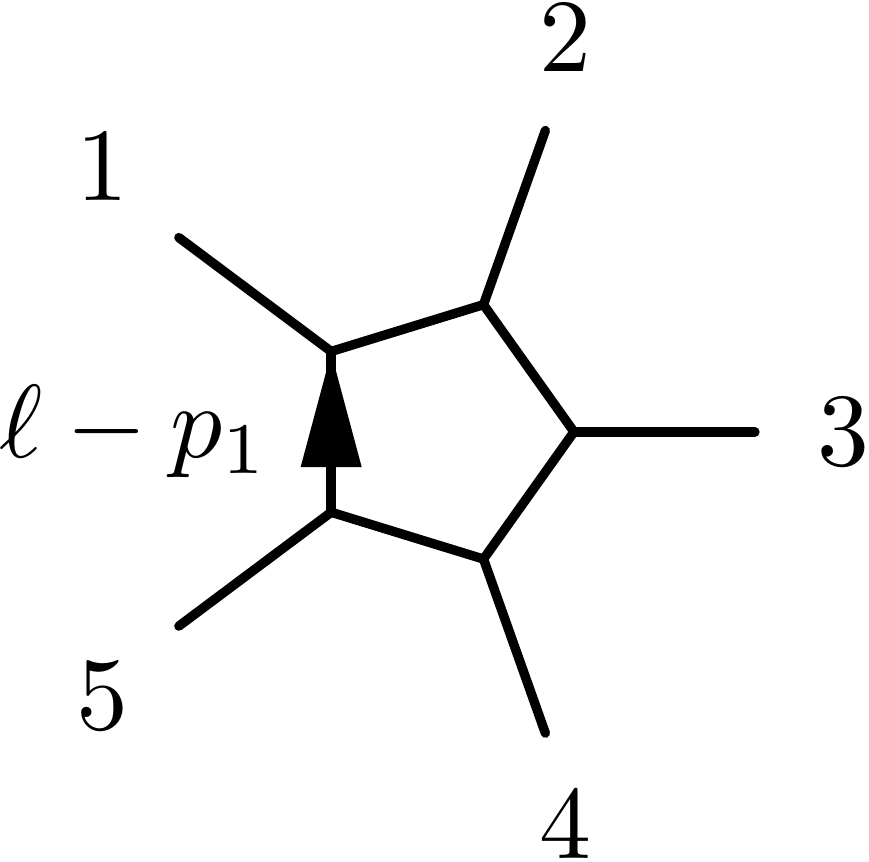}
\end{minipage} \leftrightarrow  n_5(1,2,3,4,5;\ell-p_1)\,.
\end{equation}
It follows that
\begin{equation}
n_5(1,2,3,4,5;\ell) - n_5(2,3,4,5,1;\ell) = n_{1,5}(1,2,3,4,5)\cdot p_1 \, .
\end{equation}
Using this observation in \eqn{eq:5ptwalking}, we discover a simple expression for the linear pentagon numerator, $n_{1,5}$:
\begin{equation}
n_4([1,2],3,4,5) + n_4(2,[1,3],4,5) + n_4(2,3,[1,4],5) + n_4(2,3,4,[1,5]) = n_{1,5}(1,2,3,4,5)\cdot p_1\,.
\end{equation}
Now, using the Carrasco-Johansson box numerator, \eqn{eq:5ptboxnum}, we find that the left-hand-side of this expression vanishes. This is consistent with the fact that the Carrasco-Johansson numerators are independent of loop momentum, so that $n_{5}(1,2,3,4,5;\ell)=n_{0,5}(1,2,3,4,5)$. At higher points, we will see below that dual numerators do depend on loop momentum. Moreover, relations of this type, which fix the tensor numerators in terms of the scalar numerators, generalise straightforwardly. Thus, knowledge of the scalar numerators is sufficient to recover the full $\ell$ dependence of the numerators. This is a key step in our oxidation method.

Another useful equation can be found in a similar manner, by using the reflection symmetry of the pentagon,
\begin{equation}
n_5(1,2,3,4,5) = -n_5(5,4,3,2,1)\,,
\end{equation}
where we no longer consider loop momentum dependence. The sign follows from the antisymmetric cubic vertices. As before, we begin by moving leg $1$ until it sits beside leg $5$:
\begin{align}
n_4([1,2],3,4,5) &= n_5(1,2,3,4,5) - n_5(2,1,3,4,5) \,,\nn\\
n_4(2,[1,3],4,5) &= n_5(2,1,3,4,5) - n_5(2,3,1,4,5) \,,\nn\\
n_4(2,3,[1,4],5) &= n_5(2,3,1,4,5) - n_5(2,3,4,1,5) \,.
\end{align}
Our goal now is to move the other legs until we reverse the order of all the legs:
\begin{align}
n_4([2,3],4,1,5) &= n_5(2,3,4,1,5) - n_5(3,2,4,1,5) \,,\nn\\
n_4(3,[2,4],1,5) &= n_5(3,2,4,1,5) - n_5(3,4,2,1,5) \,,\nn\\
n_4([3,4],2,1,5) &= n_5(3,4,2,1,5) - n_5(4,3,2,1,5) \nonumber \\
&= n_5(3,4,2,1,5) - n_5(5,4,3,2,1) \,.
\end{align}
The summation telescopes to
\begin{align}
2 n_5(1,2,3,4,5) &= n_5(1,2,3,4,5) - n_5(5,4,3,2,1) \nn\\
&= n_4([1,2],3,4,5) + n_4(2,[1,3],4,5) + n_4(2,3,[1,4],5) \nn\\
&\phantom{=} + n_4([2,3],4,1,5) + n_4(3,[2,4],1,5) + n_4([3,4],2,1,5) \nonumber \\
&= -2i \delta^8(Q) \frac{\sqb{12}\sqb{23}\sqb{34}\sqb{45}\sqb{51}}{\trf{1234}}\,.
\end{align}
Thus, knowledge of the BCJ box numerators is sufficient to reconstruct the BCJ pentagon numerator. This was already explained in \cite{Carrasco:2011mn} in the five-point case. In the next section, we will give an explicit construction of the BCJ box numerators for MHV amplitudes. For non-MHV amplitudes, however, another approach will be necessary, and we will return to this question in \sect{NMHVSection}.


\section{Self-dual Boxes in MHV Amplitudes}
\label{SelfDualSection}

In this section, we present a set of BCJ numerators for one-loop box diagrams in $\mN=4$ MHV amplitudes. There are two ingredients. The first is a relationship between one-loop amplitudes in self-dual gauge theory and in the MHV sector of the $\mN=4$ theory~\cite{Bern:1996ja}. The second is the evaluation of the self-dual amplitudes using $D$-dimensional unitarity cuts~\cite{Badger:2008cm}. Together, they allow for the construction of a linear map between BCJ box numerators and box-cut contributions to the self-dual amplitudes.

Let us start with the remarkable relationship of ref.~\cite{Bern:1996ja}. Denoting by $A_{n;1}$ the one-loop leading-colour partial amplitudes, it can be expressed as
\begin{equation}
\label{eq:SDMHV}
\delta^8(Q) A_{n;1}^{\mN=0}(1^+,2^+,\ldots,n^+) = -2 \epsilon (1-\epsilon) (4\pi)^2 \left[ A_{n;1}^{\mN=4}(1,2,\ldots,n)  \right]_{D\to D+4}\,,
\end{equation}
where $D=4-2\epsilon$. On the left-hand-side, we have the one-loop amplitude in pure Yang-Mills theory with all-plus external helicities, which corresponds to the self-dual sector of the theory~\cite{Cangemi:1996rx,Chalmers:1996rq}. The dimension shift $D\to D+4$ of the $\mN=4$ amplitude is to be taken before integration, when the amplitude is expressed in terms of $D$-dimensional scalar integrals $I_m^D$ with $m$ external particles, generically massive:
\begin{equation}
I_m^D \equiv \int \frac{d^D\ell}{(2\pi)^{D}}   \frac{1}{\ell^2(\ell+p_1)^2\cdots(\ell-p_{m})^2}\,.
\end{equation}
The shift means that we substitute $\epsilon \to \epsilon -2$ so that $I_m^D \to I_m^{D+4}$.

Although the expression above was conjectured to hold to all orders in $\epsilon$, we will only make use of the $\epsilon \to 0$ limit. In this limit, we have $D \to 8$ as the dimension shift, and the right-hand-side of \eqn{eq:SDMHV} is non-vanishing because the box integrals all diverge in the ultraviolet as $(4\pi)^4I^{8-2\epsilon}_4 \sim i/6\epsilon$. On the other hand, higher-degree scalar integrals (pentagons, hexagons, etc.) are ultraviolet finite in eight dimensions, and therefore will not contribute when $\epsilon \to 0$.\footnote{Even if pentagons, hexagons, etc., have a tensorial structure, the  dimension shift still works if this structure does not lead to a pole in $\epsilon$. For instance, a pentagon integral whose numerator depends linearly on loop momentum does not contribute.}
Hence we find
\begin{equation}
\label{eq:SDMHV2}
\delta^8(Q) A_{n;1}^{\mN=0}(1^+,2^+,\ldots,n^+) = -\frac{i}{3(4\pi)^2} \sum_{\textrm{boxes}\; I_4} A_{n;1}^{\mN=4}(1,2,\ldots,n) \Big|_{I_4}\,,
\end{equation}
where the sum is over the kinematic coefficients of the scalar box integrals.

We would now like to decompose the self-dual amplitude into an analogous sum over box contributions, so that we have the chance to identify the elements in that sum with the kinematic coefficients of the box integrals in the $\mN=4$ amplitude. Such a decomposition of the self-dual amplitude is exactly what the unitarity method of ref.~\cite{Badger:2008cm} provides.

The self-dual amplitude, which is the one-loop amplitude in pure Yang-Mills theory with all-plus external helicities, has vanishing four-dimensional unitarity cuts. However, it can be constructed from $D$-dimensional box cuts, as shown in \cite{Badger:2008cm},
\begin{equation}
\label{badgerA}
(4\pi)^2 A_{n;1}^{\mN=0}(1^+,2^+,\ldots,n^+) = 
-\frac{1}{24}\sum_{i=1}^{n} \sum_{j=i+1}^{i-3} \sum_{k=j+1}^{i-2} \sum_{l=k+1}^{i-1} C_{i+1,\ldots, j|j+1,\ldots, k|k+1,\ldots, l|l+1,\ldots, i} \,,
\end{equation}
where we identify $n+1 \sim 1$ and take $\epsilon \to 0$. Each box cut $C$ is computed in the same manner as a four-dimensional maximal cut, except that the four-dimensional loop momentum is infinitely massive, $\ell^2=-\ell_{(-2\epsilon)}^2\equiv\mu^2 \to \infty$, instead of being massless. In this limit of large $\mu$, the two cut solutions are given by
\begin{equation}
\label{eq:llmu}
{\ell^\pm} = \pm \mu\, \bar{\ell} + {\mathcal O}(\mu^0)\,, \qquad \textrm{with} \qquad
\bar{\ell}^\lambda = \frac{\omega^\lambda}{\left( \omega \cdot \omega \right)^{1/2}}\,,
\quad \omega^\lambda= \epsilon^{\lambda \nu\rho\sigma} K_{1\nu}K_{2\rho}K_{3\sigma}\,,
\end{equation}
where we partitioned the external momenta into the four corners of the box cut $C$, so that $K_1=p_{i+1} + \ldots + p_j$, and so on. 
Notice that, because of the Levi-Civita symbol in $\omega^\lambda$, if we permute the corners of the box the momentum $\bar{\ell}$ will be the same up to a change of sign, which will not affect the cut result. The contribution from a box cut $C$ is
\begin{equation}
\label{badgerC}
C_{i+1,\ldots, j|j+1,\ldots, k|k+1,\ldots, l|l+1,\ldots, i}=2i \lim_{\mu \to \infty} \frac{1}{\mu^4} \bar{A}_1\,\bar{A}_2\,\bar{A}_3\,\bar{A}_4  \big|_{\ell=\mu\, \bar{\ell}}\,,
\end{equation}
where the $\bar{A}_r$ are the subamplitudes in each corner of the box. For instance, $\bar{A}_1$ is the subamplitude for the absorption of the external gluons $i+1$ to $j$ by a massive scalar of momentum $\mu\, \bar{\ell}$. For $\bar{A}_2$, we have the absorption of the gluons $j+1$ to $k$ by the same massive scalar, and so on; the mass of the scalar can always be taken as $\mu$ since the corrections from the gluons' momenta is ${\mathcal O}(\mu^0)$. The leading behaviour of each $\bar{A}_r$ is linear in $\mu$. We need these tree-level amplitudes up to six points  for the cuts that we consider in this paper; they can be found in refs.~\cite{Bern:1995db,Bern:1996ja,Badger:2005zh,Badger:2008cm}, and an all-multiplicity formula is given in ref.~\cite{Forde:2005ue}.

At this stage, we can already make an observation that will be crucial later on. Each box cut is symmetric under permutations of its corners, provided the order of the external gluons in each corner is preserved. That is,
\begin{equation}
\label{eq:Csym}
C_{i+1,\ldots, j|j+1,\ldots, k|k+1,\ldots, l|l+1,\ldots, i}=C_{j+1,\ldots, k|i+1,\ldots, j|k+1,\ldots, l|l+1,\ldots, i}\,.
\end{equation}
The importance of this fact is that BCJ box numerators should also be symmetric under the permutation of its corners; otherwise, the Jacobi relations will require the presence of BCJ triangle numerators, which (we assume) are not required for $\mN=4$ amplitudes. Let us also mention another useful property:
\begin{equation}
\label{eq:Cinv}
C_{i+1,\ldots, j|j+1,\ldots, k|k+1,\ldots, l|l+1,\ldots, i}=(-1)^{1+\#(i+1,\ldots, j)} C_{\textrm{inv}(i+1,\ldots, j)|j+1,\ldots, k|k+1,\ldots, l|l+1,\ldots, i}\,,
\end{equation}
where $\textrm{inv}(123\cdots m)$ is the inverse ordering $m\cdots 321$, and $\#$ gives the number of external gluons in that corner (then $\#$ counts the three-point vertices which get inverted in $\bar{A}_r$, while the other minus sign comes from the flip $\ell \to -\ell$).
We will now give some examples, up to seven-point amplitudes. The six- and seven-point cases already illustrate all the issues involved at higher points.

\subsection*{Four Points}

In the four-point case, we get
\begin{equation}
(4\pi)^2 A_{4;1}^{\mN=0}(1^+,2^+,3^+,4^+) = - \frac{1}{6} C_{1|2|3|4} = -\frac{i}{3} \frac{\sqb{12}\sqb{34}}{\ab{12}\ab{34}}\,.
\end{equation}
As is well known, this last expression does not depend on the ordering of the particles, so that \eqn{eq:Csym} is satisfied. It follows from the dimension-shifting formula \eqn{eq:SDMHV2} that the box contributions to the $D$-dimensional $\mN=4$ amplitude give
\begin{equation}
\mA^{\mN=4}_{4}= i g^4 n_4(1,2,3,4) \left(c_{1,2,3,4} I_4^{1|2|3|4}+c_{1,3,4,2} I_4^{1|3|4|2}+c_{1,4,2,3} I_4^{1|4|2|3}\right)\,,
\end{equation}
with
\begin{equation}
 n_4(1,2,3,4) =-\frac{1}{2} \delta^8(Q)  C_{1|2|3|4} = i \delta^8(Q) \frac{\sqb{12}\sqb{34}}{\ab{12}\ab{34}}\,,
\end{equation}
where we defined
\begin{equation}
c_{1,2,3,4}=\tf^{b_1a_1b_2}\tf^{b_2a_2b_3}\tf^{b_3a_3b_4}\tf^{b_4a_4b_1}\,,
\end{equation}
and $I_4^{1|2|3|4}$ is the box integral with the corresponding propagator structure.
This well-known expression is already the full four-point MHV amplitude. To confirm that it is in BCJ form, we need to check that the Jacobi identities hold. The only type of Jacobi identity is the one connecting BCJ boxes to BCJ triangles:
\begin{equation}
n_4(1,2,3,4)-n_4(2,1,3,4) = n_3([1,2],3,4)\,.
\end{equation}
In our case, $n_4(1,2,3,4)$ is permutation symmetric. Therefore, the BCJ triangles vanish. It will still be true at higher points that the BCJ boxes are the lowest-order polygon required, as a consequence of the property \eqref{eq:Csym}.

\subsection*{Five Points}

In the five-point case, we have
\begin{equation}
(4\pi)^2 A_{5;1}^{\mN=0}(1^+,2^+,3^+,4^+,5^+) = - \frac{1}{6} \left(C_{12|3|4|5}+C_{23|4|5|1}+C_{34|5|1|2}+C_{45|1|2|3}+C_{51|2|3|4}\right)\,,
\end{equation}
where
\begin{equation}
 C_{12|3|4|5} = 2i \frac{\sqb{12}\sqb{34}\sqb{45}\sqb{53}}{\ab{12}\trf{1234}}\,,
\end{equation}
and similarly for the cyclic permutations.

We see now how useful the decomposition \eqref{badgerA} is; we can easily recognise to which box integral these different cuts correspond. The BCJ box contributions to the $\mN=4$ amplitude are then
\begin{equation}
\label{eq:5ptboxes}
\mA^{\mN=4}_{5}\bigg|_{\textrm{BCJ box}} = i g^5 \left(\frac{n_4([1,2],3,4,5)}{s_{12}} c_{[1,2],3,4,5} I_4^{(1{+}2)|3|4|5}+\textrm{9 perms}\right)\,,
\end{equation}
with
\begin{equation}
 n_4([1,2],3,4,5)= -\frac{1}{2} \delta^8(Q)  s_{12} C_{12|3|4|5} = i\delta^8(Q) \frac{[12]^2[34][45][53]}{\tr_5(1234)}\,.
\end{equation}
The 10 permutations in \eqn{eq:5ptboxes} correspond to different choices of the two particles in the massive corner of the box. 
We also defined
\begin{equation}
c_{[1,2],3,4,5}=\tf^{a_1a_2b_1}\tf^{b_5b_1b_3}\tf^{b_2a_4b_3}\tf^{b_3a_5b_4}\tf^{b_4a_6b_5}\,,
\end{equation}
and
\begin{equation}
I_4^{(1{+}2)|3|4|5} = \int \frac{d^D\ell}{(2\pi)^{D}}   \frac{1}{\ell^2(\ell+p_{12})^2(\ell-p_{45})^2(\ell-p_5)^2}\,.
\end{equation}
This set of BCJ boxes coincides with the Carrasco-Johansson representation we discussed in \sect{FivePointEasySection}.

\subsection*{Six Points}

The six-point case requires for the first time an additional ingredient. Consider\footnote{The relative factors ensure that there is no over-count for the box $C_{12|3|45|6}=C_{45|6|12|3}$.}
\begin{equation}
(4\pi)^2 A_{6;1}^{\mN=0}(1^+,2^+,3^+,4^+,5^+,6^+) = - \frac{1}{12} \left(2C_{123|4|5|6}+2C_{12|34|5|6}+C_{12|3|45|6}+ \textrm{cyclic}\right)\,,
\end{equation}
where
\begin{align}
\label{eq:sixptselfdualboxes}
C_{123|4|5|6} &= 2i \frac{\big( s_{45}\langle6|1{+}2|3]\, \sqb{51}\,\sqb{64}^2 - s_{46}\langle5|1{+}2|3]\,\sqb{61}\,\sqb{54}^2 \big) \sqb{56}}{\ab{12}\ab{23} \trf{1456}\trf{3456}}\,, \nn \\
C_{12|34|5|6} &= -2i\frac{\langle5|1{+}2|6]\, \langle6|1{+}2|5]\, \sqb{12}\,\sqb{34}\,\sqb{56}^2}{\ab{12}\ab{34} \trf{1256}\trf{3456}}\,, \nn \\
C_{12|3|45|6}&=C_{12|45|3|6}\,.
\end{align}
There are three different cubic diagrams for the one-mass boxes with particles 1,2,3 in the massive corner, such as $C_{123|4|5|6}$ . The numerators of those diagrams are
\begin{equation}
n_4([[1,2],3],4,5,6), \qquad n_4([[2,3],1],4,5,6), \qquad n_4([[3,1],2],4,5,6)\,.
\end{equation}
In terms of these numerators, we have
\begin{align}
-\frac{1}{2} \delta^8(Q) C_{123|4|5|6}&= \frac{n_4([[1,2],3],4,5,6)}{s_{12}s_{123}} - \frac{n_4([[2,3],1],4,5,6)}{s_{23}s_{123}}\,, \nn\\
-\frac{1}{2} \delta^8(Q) C_{213|4|5|6}&=-\frac{n_4([[1,2],3],4,5,6)}{s_{12}s_{123}} + \frac{n_4([[3,1],2],4,5,6)}{s_{13}s_{123}}\,, \nn\\
-\frac{1}{2} \delta^8(Q) C_{231|4|5|6}&= \frac{n_4([[2,3],1],4,5,6)}{s_{23}s_{123}} - \frac{n_4([[3,1],2],4,5,6)}{s_{13}s_{123}}\,.
\end{align}
Only two of these relations are independent, and one can verify the identity
\begin{align}
C_{123|4|5|6}+C_{213|4|5|6}+C_{231|4|5|6}=0\,,
\end{align}
using the explicit expressions in \eqn{eq:sixptselfdualboxes}. Therefore, we need an additional relation in order to fix the three BCJ box numerators in terms of the box cuts. This relation is naturally the Jacobi identity:
\begin{equation}
n_4([[1,2],3],4,5,6) + n_4([[2,3],1],4,5,6) + n_4([[3,1],2],4,5,6) =0\,.
\end{equation}
We may now solve for the numerators in terms of the cut boxes, and find
\begin{equation}
n_4([[1,2],3],4,5,6) = -\frac{1}{2} \delta^8(Q) s_{12}\left(s_{23} C_{123|4|5|6}- s_{13} C_{213|4|5|6} \right)\,.
\end{equation}

The BCJ box numerators corresponding to the two-mass boxes can be obtained in a straightforward manner,
\begin{equation}
 n_4([1,2],[3,4],5,6)= -\frac{1}{2} \delta^8(Q) s_{12}s_{34} C_{12|34|5|6}\,.
\end{equation}
Notice that, again due to the property \eqref{eq:Csym} of the cut boxes, there is no difference between the so-called two-mass-hard (massive corners opposed) and two-mass-easy (massive corners adjacent) boxes,
\begin{equation}
 n_4([1,2],3,[4,5],6) = n_4([1,2],[4,5],3,6)\,.
\end{equation}

The BCJ box numerators obtained in this section coincide with the ones chosen by Yuan in ref.~\cite{Yuan:2012rg}; the present discussion may be regarded as a systematic derivation of these boxes. We will show below why it was not possible in ref.~\cite{Yuan:2012rg} to obtain the BCJ pentagon and hexagon numerators satisfying all the Jacobi identities: those numerators will depend on the loop momentum. Indeed, we will see that this dependence is fixed by the choice of box numerators.

\subsection*{Seven Points}

At seven points, the simplest case is that of the BCJ numerators for three-mass boxes,
\begin{equation}
n_4([1,2],[3,4],[5,6],7) = -\frac{1}{2} \delta^8(Q) s_{12}s_{34} s_{56} C_{12|34|56|7}\,.
\end{equation}
For the two-mass boxes, one has to use the Jacobi identity as in the six-point case, with the result
\begin{equation}
n_4([[1,2],3],[4,5],6,7) = -\frac{1}{2} \delta^8(Q) s_{12}s_{45} \left(  s_{23} C_{123|45|6|7}- s_{31} C_{312|45|6|7} \right)\,.
\end{equation}
For the one-mass boxes, we have two different topologies of cubic diagrams,
\begin{equation}
\label{7pegnum}
n_4([[[1,2],3],4],5,6,7) \qquad \textrm{and} \qquad n_4([[1,2],[3,4]],5,6,7)\,.
\end{equation}
If we consider all the permutions of the particles 1 to 4, these topologies give us 12 and 3 different cubic diagrams, respectively. These 15 numerators satisfy 10 Jacobi identities among themselves, but only 9 of those are independent identities. On the other hand, after considering the property \eqref{eq:Cinv}, there are 12 different corresponding box cuts, such as
\begin{align}
-\frac{1}{2} \delta^8(Q) C_{1234|5|6|7}& = \frac{n_4([[[1,2],3],4],5,6,7)}{s_{12}s_{123}s_{1234}} + \frac{n_4([[[3,2],1],4],5,6,7)}{s_{23}s_{123}s_{1234}}
 \nonumber \\
& \quad - \frac{n_4([[[2,3],4],1],5,6,7)}{s_{23}s_{234}s_{1234}} - \frac{n_4([[[4,3],2],1],5,6,7)}{s_{34}s_{234}s_{1234}}\ \nonumber \\
& \quad + \frac{n_4([[1,2],[3,4]],5,6,7)}{s_{12}s_{34}s_{1234}} \,.
\end{align}
These satisfy identities of the type
\begin{align}
C_{1234|5|6|7}+C_{2134|5|6|7}+C_{2314|5|6|7}+C_{2341|5|6|7}=0\,,
\end{align}
which reduce the number of independent box cuts to 6. Together with the 9 Jacobi identities, the independent box cuts determine the 15 BCJ box numerators. The latter can be expressed as
\begin{align}
& n_4([[[1,2],3],4],5,6,7) =  -\frac{1}{2} \delta^8(Q) s_{12} \bigg(  s_{34} \big(  s_{23} C_{1234|5|6|7} - s_{13} C_{2134|5|6|7} \big)    \nonumber\\
& \qquad \qquad + (s_{13}+s_{23}) \big( s_{14} C_{3214|5|6|7} - s_{24} C_{3124|5|6|7} \big)  + s_{13}s_{14}C_{2314|5|6|7} - s_{23}s_{24}C_{1324|5|6|7} \bigg)\,,
\end{align}
and, obviously,
\begin{align}
n_4([[1,2],[3,4]],5,6,7) = n_4([[[1,2],3],4],5,6,7) - n_4([[[1,2],4],3],5,6,7)\,.
\end{align}

\subsection*{Higher Points}

We have seen in the previous examples that the box cuts $C$ completely fix the BCJ box numerators. This will still be true for an arbitrary number of external gluons. The first step to see how this works is to notice that we can focus on one corner of the box cuts at a time. According to the expression \eqref{badgerC}, $C$ factorizes into the four corners, which are only connected through the loop momentum cut solution.

We want to identify the contribution from each corner of a box cut with a tree-level ``BCJ current," consisting of $m_r$ external gluons and one off-shell leg connected to the loop momentum. Each ``BCJ current" is a sum over diagrams with cubic vertices, whose kinematic numerators satisfy Jacobi identities. The counting of independent BCJ numerators for $m_r+1$ external legs is $(m_r-1)!$. For instance, there were 6 such numerators in the one-mass box at seven points; recall the text below \eqn{7pegnum}.

To obtain the linear map between BCJ box numerators and box cuts $C$, we need a set of $(m_r-1)!$ independent box cuts. Such a set is the result of the Kleiss-Kuijf-type relations of the type:
\begin{equation}
\label{eq:CKK}
C_{123\ldots m|\alpha|\beta|\gamma} + C_{213\ldots m|\alpha|\beta|\gamma} + C_{231\ldots m|\alpha|\beta|\gamma}+  \ldots  + C_{23\ldots 1m|\alpha|\beta|\gamma} + C_{23\ldots m1|\alpha|\beta|\gamma} =0\,,
\end{equation}
where $\alpha$, $\beta$ and $\gamma$ are the sets of external gluons in the other three corners. These relations were proven in \cite{Boels:2011tp}, where an explicit solution in terms of $(m_r-1)!$ independent box cuts was obtained,
\begin{equation}
C_{\delta 1 \delta' |\alpha|\beta|\gamma} = (-1)^{|\delta|}\sum_{\sigma \in OP(\delta^T \cup \delta')} C_{1 \sigma |\alpha|\beta|\gamma} \,,
\end{equation}
where the sum is over all permutations $\sigma$ that preserve the ordering of the elements in each subset $\delta^T$ and $\delta'$; $\delta^T$ denotes the transposition of the elements in $\delta$, and $|\delta|$ denotes the number of elements in $\delta$. It was also shown in \cite{Boels:2011tp} that these relations are equivalent to the relations between self-dual partial amplitudes found in \cite{BjerrumBohr:2011xe}. 

Let us mention the closely related work of ref. \cite{Mafra:2012kh}, where certain BRST invariant kinematic combinations found in superstring amplitudes have properties analogous to the box numerators constructed in this section. This may provide an alternative method to compute BCJ box numerators.


\section{Integrand Reduction}
\label{ReductionsSection}

In this section, we briefly discuss the problem of reducing scalar integrals algebraically at the integrand level. This will be important in order to compare the BCJ representations constructed in this paper with previously obtained representations of the amplitudes.

Consider a scalar $m$-gon integral in $D$ dimensions, given by 
\begin{equation}
 I_m^D \equiv \int \frac{d^D \ell}{(2\pi)^D} \frac{1}{\prod_{i=1}^m P_i}\,,
\end{equation}
which has massless propagators
\begin{equation}
 P_i \equiv (\ell+p_1+\ldots+p_i)^2\,,
\end{equation}
and whose external momenta $p_i$ need not be null. With all external momenta in four dimensions and $m\ge5$, we can partial fraction the integrand into a sum of scalar pentagon integrands
\begin{equation}
\label{partialfraction}
 \frac{1}{\prod_{i=1}^{m} P_i} = \sum \frac{r((j_5{+}1)\ldots j_1|(j_1{+}1)\ldots j_2|(j_2{+}1)\ldots j_3|(j_3{+}1)\ldots j_4|(j_4{+}1)\ldots j_5)}{P_{j_1}P_{j_2}P_{j_3}P_{j_4}P_{j_5}}\,,
\end{equation}
where the coefficients $r(\ldots)$ depend on five ordered partitions of the set of external legs. By definition, $r$ is cyclic in its arguments: $r(\alpha_1|\alpha_2|\alpha_3|\alpha_4|\alpha_5) =r(\alpha_2|\alpha_3|\alpha_4|\alpha_5|\alpha_1)$. Although the notation in \eqn{partialfraction} may seem needlessly lengthy, it will be crucial to be able to distinguish between different orderings of the external legs, so that, for example, $r(12|3|4|5|6)$ and $r(21|3|4|5|6)$ are not the same quantity. In particular, by $r(21|3|4|5|6)$ we mean the coefficient of $\left(\ell^2 (\ell+p_{12})^2 (\ell+p_{123})^2 (\ell+p_{1234})^2 (\ell+p_{12345})^2\right)^{-1}$ in the partial fractioning of the scalar hexagon integrand with external legs $213456$, where $p_{12} = p_1+p_2$ and so forth. We may also write something like ${r((1{+}2)3|4|5|6|7)}$, which is the coefficient of the pentagon with corners $(1{+}2{+}3,4,5,6,7)$ in the partial fractioning of a hexagon with corners $(1{+}2,3,4,5,6,7)$. 

This partial fractioning of scalar integrands was given explicitly in ref.~\cite{Melrose:1965kb} in terms of modified Cayley determinants. More recently, equivalent expressions were given in ref.~\cite{Bern:1992em}, which contains a more careful consideration of the dimensional regulator.

In the present work, we prefer to use somewhat less cumbersome expressions in terms of traces of gamma matrices. As a simple example, consider the reduction of a hexagon integral, where we have
\begin{equation}
\label{hexreduction}
 r(12|3|4|5|6) = -\frac{\trf{3456}}{\trf{123456}}\,.
\end{equation}
\eqn{hexreduction} can be proven straightforwardly by verifying the identity
\begin{align}
 1 &= r(12|3|4|5|6) P_1 + r(1|23|4|5|6) P_2 + r(1|2|34|5|6) P_3  \nn\\
  &\phantom{=} +r(1|2|3|45|6) P_4 + r(1|2|3|4|56) P_5 + r(61|2|3|4|5) P_6\,,
\label{eq:sumrP}
\end{align}
for example by considering the quantity $\trf{12345\ell}$ and anti-commuting $\ell$ around the trace.
As a more complicated example, the reduction of an octagon integral requires the quantity
\begin{equation}
 r(123|4|56|7|8) = -\frac{\trf{4(5{+}6)78}}{\trf{1(2{+}3)4(5{+}6)78}} \frac{\trf{4(5{+}6)78}}{\trf{(1{+}2)34(5{+}6)78}} 
    \frac{\trf{(1{+}2{+}3)478}}{\trf{(1{+}2{+}3)45678}}\,.
\end{equation}
The three factors in this expression correspond to the three places one can insert an extra propagator in the pentagon specified by $(123|4|56|7|8)$ while respecting the ordering of the external legs. Following these two examples, it is not difficult to guess the general expression, which follows by combining \eqn{hexreduction} with the results of ref.~\cite{Melrose:1965kb}:
\begin{equation}
 r(\alpha_1|\alpha_2|\alpha_3|\alpha_4|\alpha_5) = \prod_{j=1}^5 \prod_{\{\beta_1,\beta_2\}=\alpha_j} \left(-\frac{\trf{p_{\alpha_1}\ldots \hat{p}_{\alpha_j} \ldots p_{\alpha_5}}}{\trf{p_{\alpha_1}\ldots p_{\beta_{1}}p_{\beta_{2}}\ldots p_{\alpha_5}}}\right)\,,
\end{equation}
where $\{\beta_1,\beta_2\}=\alpha_j$ indicates all ways of splitting $\alpha_j$ into two subsets $\beta_1$ and $\beta_2$ that respect the ordering of $\alpha_j$, and the hat in the numerator indicates the absence of $p_{\alpha_j}$ from the trace. 

It is important to recognise that the reduction of a scalar $m$-gon to scalar pentagons in \eqn{partialfraction} is an algebraic identity on the integrand, relying only on having four-dimensional external momenta and not on any integration identities such as vanishing of total derivatives. This will be useful in our discussion of the BCJ form of the six- and seven-point amplitudes, whose numerators depend on loop momentum; it will allow us to perform integrand reduction while leaving the numerators unspecified.

It is also possible to reduce scalar pentagon integrals to scalar boxes, as in refs.~\cite{Bern:1992em,'tHooft:1978xw,vanNeerven:1983vr}. Unlike reduction to pentagons, however, reduction to boxes is not an algebraic identity; it throws away parity-odd integrands that integrate to zero and is only valid up to terms of $O(\epsilon)$. Nevertheless, we will find it useful to compare with known expressions for amplitudes, given as linear combinations of scalar box integrals. The reduction from scalar pentagons to scalar boxes is then achieved using the reduction coefficient
\begin{equation}
 r_4(12|3|4|5) = \frac{\tr(34512345)-4 (p_1\cdot p_2) \, p_3^2\, p_4^2\, p_5^2}{4 p_1^2\, p_2^2\, p_3^2\, p_4^2\, p_5^2-\tr(1234512345)}\,.
\end{equation}

\noindent Finally, one can explicitly reduce $m$-gons to boxes by first reducing to pentagons, yielding 
\begin{equation}
 r(\alpha_1|\alpha_2|\alpha_3|\alpha_4) = \sum_{j=1}^4 \sum_{\{\beta_1,\beta_2\}=\alpha_j} r_4(\ldots|\beta_1\beta_2|\ldots) r(\ldots|\beta_1|\beta_2|\ldots)\,.
\end{equation}
%


\section{The Six-point MHV Amplitude}
\label{SixPointSection}

We are now in a position to compute the numerators of the six-point one-loop MHV amplitude. In fact, the computation is remarkably straightforward if we follow the guidance from the five-point case. We begin with the box numerators discussed in \sect{SelfDualSection}, namely 
\begin{align}
n_4([[1,2],3],4,5,6) &= -\frac{1}{2} \delta^8(Q) s_{12}\left(  s_{23} C_{123|4|5|6}- s_{13} C_{213|4|5|6} \right)\,, \\
n_4([1,2],[3,4],5,6) &= -\frac{1}{2} \delta^8(Q) s_{12} s_{34} C_{12|34|5|6}\,.
\end{align}
These box numerators do not depend on loop momentum. More generally, we do expect the numerators to depend on loop momentum; indeed, it is reasonable to guess that the $m$-gon numerator will have up to $m-4$ powers of loop momentum.

Our first task is to determine the pentagon numerators $n_5(a,b,c,d,e;\ell)$.\footnote{Recall that the subscripted 5 refers to the pentagon, regardless of what multiplicity amplitude we are discussing.}
Since we are dealing with a six-point amplitude, one of the legs of the pentagon is massive; we will treat both massive and massless legs on the same footing.
We begin by following the integrand oxidation procedure we described in \sect{FivePointEasySection} -- namely, we move one leg around the loop. In detail, we consider these equations:
\begin{align}
\label{eq:pentperm16pt}
n_5(a,b,c,d,e;\ell) - n_5(b,a,c,d,e;\ell)  &= n_4([a,b],c,d,e)\,, \\
n_5(b,a,c,d,e;\ell) - n_5(b,c,a,d,e;\ell)  &= n_4(b, [a,c],d,e)\,, \\
&\vdots \nn\\
n_5(b,c,d,a,e;\ell) - n_5(b,c,d,e,a;\ell) &= n_4(b,c,d,[a,e])\,.
\end{align}
Notice that we have moved the leg $a$ around the loop, so we expect that $n_5(b,c,d,e,a;\ell)$ is related to $n_5(a,b,c,d,e;\ell)$. Indeed, the cyclic permutation property of the pentagon is
\begin{equation}
n_5(b,c,d,e,a;\ell) = n_5(a,b,c,d,e;\ell -p_a)\,.
\end{equation}
Therefore, by summing our system of equations, we learn that
\begin{equation}
\label{eq:directPentagon}
n_5(a,b,c,d,e;\ell) - n_5(a,b,c,d,e;\ell-p_a) = x_5(a,b,c,d,e)\,,
\end{equation}
where
\begin{equation}
 x_5(a,b,c,d,e)\equiv n_4([a,b],c,d,e)  + n_4(b,[a,c],d,e]) + n_4(b,c,[a,d],e) + n_4(b,c,d,[a,e])\,.
\end{equation}
Now, let us consider expanding $n_5(a,b,c,d,e;\ell)$ in the loop momentum $\ell$:
\begin{equation}
n_5(a,b,c,d,e;\ell) = n_{0,5}(a,b,c,d,e) + n_{1,5}(a,b,c,d,e) \cdot \ell + \ell \cdot n_{2,5}(a,b,c,d,e) \cdot \ell + \cdots
\end{equation}
The structure of \eqn{eq:directPentagon} is basically a finite difference equation. Since $x_5(a,b,c,d,e)$ is independent of loop momentum (after all, it is a sum of boxes), we see that $n_5$ can depend on at most one power of loop momentum ($n_{2,5}=0$, etc.), consistent with our expectation. We now find that
\begin{equation}
\label{eq:n15}
p_a \cdot n_{1,5}(a,b,c,d,e) = x_5(a,b,c,d,e)\,.
\end{equation} 
This is not all we know about the numerator $n_{1,5}$. From~\eqn{eq:pentperm16pt}, whose right-hand side is independent of the loop momentum, and from the vanishing of $n_{2,5}$, we conclude that $n_{1,5}$ is permutation symmetric. This is enough to determine $n_{1,5}$ by cyclically permuting \eqn{eq:n15}:
\begin{align}
p_b \cdot n_{1,5}(a,b,c,d,e) &= x_5(b,c,d,e,a)\,, \\
p_c \cdot n_{1,5}(a,b,c,d,e) &= x_5(c,d,e,a,b)\,, \\
p_d \cdot n_{1,5}(a,b,c,d,e) &= x_5(d,e,a,b,c)\,.
\end{align}
We can therefore reconstruct the numerator as
\begin{align}
n_{1,5}^\mu(a,b,c,d,e) &= \omega_a^\mu\: x_5(a,b,c,d,e) + \omega_b^\mu\: x_5(b,c,d,e,a) \nonumber \\ 
& \quad+ \omega_c^\mu\: x_5(c,d,e,a,b) + \omega_d^\mu\: x_5(d,e,a,b,c),  
\end{align}
where
\begin{equation}
\omega_a^\mu = \frac{\epsilon(\cdot,b,c,d)}{\epsilon(a,b,c,d)}\,, \qquad \omega_b^\mu = \frac{\epsilon(a,\cdot,c,d)}{\epsilon(a,b,c,d)}\,, \qquad \textrm{etc.}                                                                                                                    
\end{equation}
The vectors $\{ \omega_a, \omega_b, \omega_c, \omega_d\}$ form a four-dimensional basis dual to $\{p_a,p_b,p_c,p_d\}$.

To determine the part of the pentagon numerator that is independent of loop momentum, we use Jacobi moves to rearrange the legs from a starting position to its reverse. Then we can use the mirror symmetry of the pentagon to deduce a simple equation. From the Jacobi moves, we deduce that
\begin{equation}
n_5(a,b,c,d,e; \ell) - n_5(c,b,a,e,d;\ell) = y_5(a,b,c,d,e) \,,
\end{equation}
where
\begin{equation}
 y_5(a,b,c,d,e) \equiv n_4([a,b],c,d,e) + n_4(b,[a,c],d,e) + n_4([b,c],a,d,e) + n_4(c,b,a,[d,e])\,.
\end{equation}
The symmetries of the pentagon require that
\begin{align}
n_5(c,b,a,e,d;\ell) &\leftrightarrow
\begin{minipage}[c]{0.2\linewidth}
\centering
\includegraphics[scale=0.3]{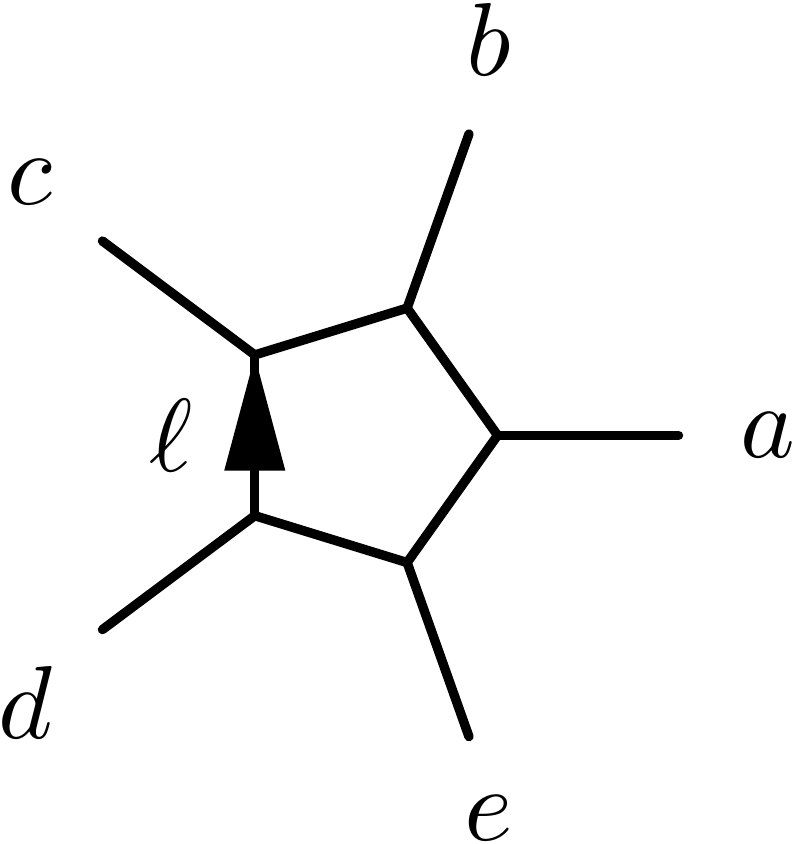}
\end{minipage} \nn\\
&=- \left(
\begin{minipage}[c]{0.2\linewidth}
\centering
\includegraphics[scale=0.3]{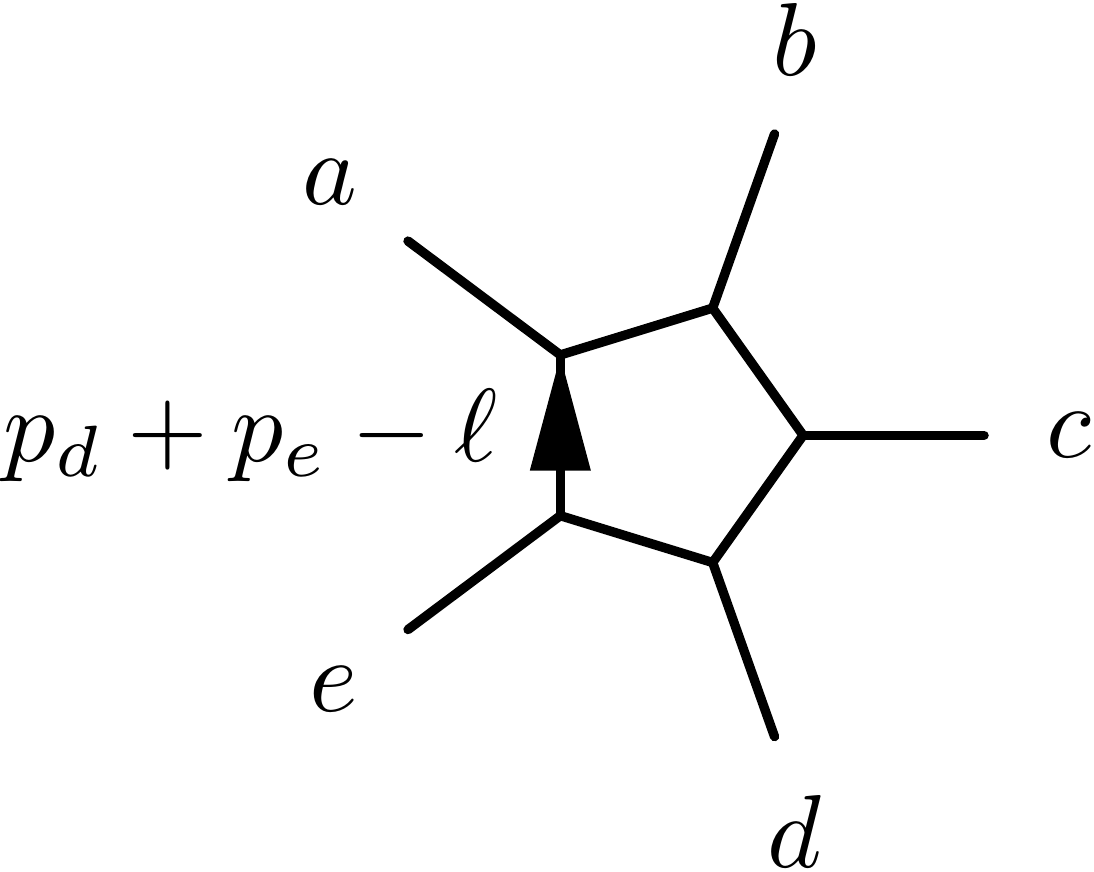}
\end{minipage}\;\; \right) \leftrightarrow
 - n_5(a,b,c,d,e; -\ell +p_d +p_e)\,,
\end{align}
allowing us to determine $n_{0,5}$ in terms of the boxes we constructed from the self-dual theory and the linear pentagon produced with the oxidation procedure,
\begin{equation}
n_{0,5}(a,b,c,d,e) = \frac 12 \left( y_5(a,b,c,d,e) - (p_d + p_e) \cdot n_{1,5}(a,b,c,d,e)\right)\,.
\end{equation}

We have now fully constructed the pentagon numerator. It is straightforward to construct the hexagon numerator using the same methods. First, we move one leg around the loop and use cyclicity to discover that
\begin{equation}
\label{eq:directHexagon}
n_6(a,b,c,d,e,f;\ell) - n_6(a,b,c,d,e,f;\ell-p_a) = x_6(a,b,c,d,e;\ell)\,, 
\end{equation}
where
\begin{align}
\label{eq:x6}
 x_6(a,b,c,d,e;\ell) & \equiv  n_5([a,b],c,d,e,f;\ell) + n_5(b,[a,c],d,e,f;\ell) + n_5(b,c,[a,d],e,f;\ell) \nn\\
  &\phantom{\equiv} + n_5(b,c,d,[a,e],f;\ell) + n_5(b,c,d,e,[a,f];\ell)\,.
\end{align}
Notice that $x_6$, which is a sum of pentagons, may now depend on the loop momentum. Since there is at most one power of $\ell$ in $x_6$, the hexagon numerator can have at most two powers of $\ell$. However, a remarkable cancellation takes place. The linear combination of pentagons in $x_6$ is such that it does not depend on the loop momentum after all, a fact which boils down to our choice of box numerators. 
Therefore, the hexagon numerator depends on only one power of loop momentum,
\begin{equation}
n_6(a,b,c,d,e,f;\ell) = n_{0,6}(a,b,c,d,e,f) + n_{1,6}(a,b,c,d,e,f) \cdot \ell\,.
\end{equation}
Using this expansion in \eqn{eq:directHexagon}, we find an equivalent expression involving $n_{1,6}$ alone:
\begin{equation}
p_a \cdot n_{1,6}(a,b,c,d,e,f) = x_6(a,b,c,d,e,f)\,.
\end{equation}
Using cyclicity of $n_{1,6}$ it is straightforward to find a solution:
\begin{align}
n_{1,6}^\mu(a,b,c,d,e,f) &=  \omega_a^\mu \: x_6(a,b,c,d,e,f) + \omega_b^\mu \: x_6(b,c,d,e,f,a) \nonumber \\ 
& \quad + \omega_c^\mu \: x_6(c,d,e,f,a,b) + \omega_d^\mu \: x_6(d,e,f,a,b,c)\,.
\label{eq:n16sol1}
\end{align}
Another non-trivial cancellation occurs when we perform an integrand reduction to pentagons, if we now think about a partial amplitude. The linear loop momentum dependence of the hexagon numerator cancels against that of the pentagon numerators:
\begin{equation}
n_{1,6}(a,b,c,d,e,f)\: r(ab|c|d|e|f) + \frac{n_{1,5}([a,b],c,d,e,f)}{s_{ab}} =0\,.
\label{eq:n16sol2}
\end{equation}
The consistency between the expressions \eqref{eq:n16sol1} and \eqref{eq:n16sol2} shows the tight connection between the system of Jacobi identities and the structure of integrand reductions.

It remains to determine $n_{0,6}$. In the case of the pentagon, we used inversion symmetry to find the analogous part of the numerator. But for the hexagon this is no longer of any help. The reason is simply that $n_6(a,b,c,d,e,f;\ell) = + n_6(f,e,d,c,b,a;-\ell)$ since we are now reversing the orientation of six antisymmetric cubic vertices. Thus, the inversion symmetry leads to an equation of the form
\begin{equation}
2 n_{1,6}(a,b,c,d,e,f)\cdot \ell = y_6(a,b,c,d,e,f;\ell)\,,
\end{equation}
where $y_6$ is a linear combination of pentagons. In fact, the procedure we have been following of systematically exploiting the Jacobi relations and the symmetries of the loop diagrams cannot fix the scalar part of a hexagon numerator. To see this, suppose we have the correct scalar hexagon numerator $n_{0,6}$ in our hands. Consider adding a function $S(a,b,c,d,e,f)$ of kinematic variables to this numerator:
\begin{equation}
\tilde n_{0,6}(a,b,c,d,e,f) = n_{0,6}(a,b,c,d,e,f) + S(a,b,c,d,e,f)\,.
\end{equation}
The Jacobi relations, and the cyclicity and inversion symmetries of the hexagon, are unchanged provided that $S$ is a permutation-symmetric function. To fix this ambiguity we must add more physical information. We choose to do so simply by requiring that our numerators reproduce the known MHV amplitude \cite{Bern:1994zx,Bern:1998sv}. We use integrand reduction to boxes; in particular, it is known that the two-mass-hard box coefficient vanishes. The reduction is straightforward, with the result that
\begin{multline}
n_{0,6}(a,b,c,d,e,f)\: r(ab|cd|e|f) + \frac{n_{0,5}([a,b],c,d,e,f)}{s_{ab}} r(a{+}b|cd|e|f)\\ + \frac{n_{0,5}(a,b,[c,d],e,f)}{s_{cd}} r(ab|c{+}d|e|f) + \frac{n_4([a,b],[c,d],e,f)}{s_{ab}s_{cd}} = 0\,.
\end{multline}
This uniquely fixes the scalar part of the hexagon numerator. The Jacobi relation to the scalar pentagon is satisfied.

In this way, we have built a set of numerators which satisfies all the Jacobi relations. In the construction, we used the absence of the two-mass-hard boxes. To verify our procedure, we numerically checked that the two-mass-easy and one-mass box coefficients take their known values.


\section{The Seven-point MHV Amplitude}
\label{SevenPointSection}

Essentially the same techniques suffice to compute the seven-point, one-loop, MHV amplitude. A new element to emerge in this case is that there is a family of numerators for the heptagon with the property that the associated boxes are those computed from the self-dual boxes. In other words, there are generalised gauge transformations among colour-dual numerators at seven points that preserve the box numerators. 

The computation starts with the boxes. There are three kinds of box numerators relevant in this computation; namely, the one-, two- and three-mass boxes. As we discussed in \sect{SelfDualSection}, we may take the numerators of these boxes to be
\begin{align}
n_4([1,2],[3,4],[5,6],7) &= -\frac{1}{2} \delta^8(Q) s_{12}s_{34} s_{56} C_{12|34|56|7} \,, \\
n_4([[1,2],3],[4,5],6,7) &= -\frac{1}{2} \delta^8(Q) s_{12}s_{45} \left(  s_{23} C_{123|45|6|7}- s_{31} C_{312|45|6|7} \right)\,,
\end{align}
for the three- and the two-mass boxes, respectively. All one-mass boxes can be computed from
\begin{align}
& n_4([[[1,2],3],4],5,6,7)=  -\frac{1}{2} \delta^8(Q) s_{12} \bigg(  s_{34} \big(  s_{23} C_{1234|5|6|7} - s_{13} C_{2134|5|6|7} \big)    \nonumber\\
& \qquad \qquad + (s_{13}+s_{23}) \big( s_{14} C_{3214|5|6|7} - s_{24} C_{3124|5|6|7} \big)  + s_{13}s_{14}C_{2314|5|6|7} - s_{23}s_{24}C_{1324|5|6|7} \bigg) \,.
\end{align}
Given these boxes, we can construct pentagon numerators in complete analogy with the hexagon case. Thus, the pentagon numerator,
\begin{equation}
  n_5(a,b,c,d,e;\ell) = n_{0,5}(a,b,c,d,e) + n_{1,5}(a,b,c,d,e) \cdot \ell\,,                                                                                                                   
\end{equation}
is given by
\begin{align}
n_{1,5}^\mu(a,b,c,d,e) &= \omega_a^\mu\: x_5(a,b,c,d,e) + \omega_b^\mu\: x_5(b,c,d,e,a) \nonumber \\ 
& \quad+ \omega_c^\mu\: x_5(c,d,e,a,b) + \omega_d^\mu\: x_5(d,e,a,b,c)\,,  \\
2n_{0,5}(a,b,c,d,e) &= y_5(a,b,c,d,e) - p_{de} \cdot n_{1,5}(a,b,c,d,e)\,,
\end{align}
where
\begin{align}
x_5(a,b,c,d,e) &\equiv n_4([a,b],c,d,e)  + n_4(b,[a,c],d,e) + n_4(b,c,[a,d],e) + n_4(b,c,d,[a,e]) \,, \\
y_5(a,b,c,d,e) &\equiv n_4([a,b],c,d,e) + n_4(b,[a,c],d,e) + n_4([b,c],a,d,e)  + n_4(c,b,a,[d,e]) \,,
\end{align}
and, just as in \sect{SixPointSection}, we define
\begin{equation}
\omega_a^\mu = \frac{\epsilon(\cdot,b,c,d)}{\epsilon(a,b,c,d)}\,, \qquad \omega_b^\mu = \frac{\epsilon(a,\cdot,c,d)}{\epsilon(a,b,c,d)}\,, \qquad \textrm{etc.}                                                                                                                    
\end{equation}

In the case of the six-point amplitude, the hexagon numerator contained at most one power of the loop momentum. This is no longer the case at seven points. By walking leg $a$ around the hexagon, we find that the hexagon numerator 
\begin{equation}
 n_6(a,b,c,d,e,f;\ell) = n_{0,6}(a,b,c,d,e,f) + n_{1,6}(a,b,c,d,e,f) \cdot \ell + \ell \cdot n_{2,6}(a,b,c,d,e,f) \cdot \ell\,
\end{equation}
satisfies
\begin{align}
2 p_{a\mu} \; n_{2,6}^{\mu\nu} (a,b,c,d,e,f) &= x_{1,6}^\nu(a,b,c,d,e,f) \,, \\
p_a \cdot n_{1,6} (a,b,c,d,e,f) &= x_{0,6} (a,b,c,d,e,f) + p_a \cdot n_{2,6}(a,b,c,d,e,f) \cdot p_a\,,
\end{align}
where $x_6(a,b,c,d,e,f; \ell ) = x_{0,6} (a,b,c,d,e,f)  + x_{1,6}(a,b,c,d,e,f) \cdot \ell$ is a sum of pentagons obtained in the integrand oxidation algorithm. In detail,
\begin{multline}
x_6(a,b,c,d,e,f; \ell ) \equiv n_5([a,b],c,d,e,f;\ell) + n_5(b,[a,c],d,e,f;\ell) \\ + n_5(b,c,[a,d],e,f;\ell) + n_5(b,c,d,[a,e],f;\ell) + n_5(b,c,d,e, [a,f];\ell)\,.
\end{multline}
The key difference relative to the six-point construction is that, in contrast to the discussion below \eqn{eq:x6}, in this case the quantity $x_{1,6}(a,b,c,d,e,f)$ does not vanish. 

It is straightforward to solve for $n_{2,6}(a,b,c,d,e,f)$. Notice that, without loss of generality, we can take $n_{2,6}$ to be a symmetric tensor. Moreover, since the pentagon numerator is at most linear in $\ell$, all the Jacobi relations involving $n_{2,6}$ must be trivial. Therefore, $n_{2,6}(a,b,c,d,e,f)$ is a permutation symmetric function of the legs $a,b,c,d,e$ and $f$. A symmetric tensor is determined if we know its projections onto a basis of four vectors, which we take to be $p_a$, $p_b$, $p_c$ and $p_d$. (Note that these momenta may be off-shell, corresponding to a massive leg.) Thus, we need to construct a tensor satisfying
\begin{align}
\label{eq:7pt_n26eqs1}
2 p_a \cdot n_{2,6} (a,b,c,d,e,f) &= x_{1,6}(a,b,c,d,e,f) \,, \\
\label{eq:7pt_n26eqs2}
2 p_b \cdot n_{2,6} (a,b,c,d,e,f) &= x_{1,6}(b,c,d,e,f,a) \,, \\
2 p_c \cdot n_{2,6} (a,b,c,d,e,f) &= x_{1,6}(c,d,e,f,a,b) \,, \\
2 p_d \cdot n_{2,6} (a,b,c,d,e,f) &= x_{1,6}(d,e,f,a,b,c) \,.
\end{align}
It is non-trivial that these equations are consistent with the symmetry of $n_{2,6}^{\mu \nu}$. Using \eqn{eq:7pt_n26eqs1} and \eqn{eq:7pt_n26eqs2}, we see that
\begin{align}
2 p_a \cdot n_{2,6}(a,b,c,d,e,f) \cdot p_b &= p_b \cdot x_{1,6}(a,b,c,d,e,f)\,, \\
2 p_b \cdot n_{2,6}(b,c,d,e,f,a) \cdot p_a &= p_a \cdot x_{1,6}(b,c,d,e,f,a)\,. 
\end{align}
Therefore consistency requires
\begin{equation}
p_b \cdot x_{1,6}(a,b,c,d,e,f) = p_a \cdot x_{1,6}(b,c,d,e,f,a) \,.
\end{equation}
We have verified this equation using the explicit self-dual boxes. In view of this consistency, the solution for the numerator $n_{2,6}^{\mu \nu}$ is simply
\begin{align}
n_{2,6}^{\mu \nu}&(a,b,c,d,e,f)= \omega_a^\mu \; \omega_a^\nu \; p_a \cdot x_{1,6}(a,b,c,d,e,f) + \omega_b^\mu \; \omega_b^\nu \; p_b \cdot x_{1,6}(b,c,d,e,f,a) \nonumber \\
& \quad + \omega_c^\mu \; \omega_c^\nu \; p_c \cdot x_{1,6}(c,d,e,f,a,b) + \omega_d^\mu \; \omega_d^\nu \; p_d \cdot x_{1,6}(d,e,f,a,b,c) \nonumber \\
& \quad + (\omega_a^\mu \omega_b^\nu + \omega_b^\mu \omega_a^\nu)  \; p_b \cdot x_{1,6}(a,b,c,d,e,f) + (\omega_a^\mu \omega_c^\nu + \omega_c^\mu \omega_a^\nu) \; p_c \cdot x_{1,6}(a,b,c,d,e,f) \nonumber\\
& \quad + (\omega_a^\mu \omega_d^\nu + \omega_d^\mu \omega_a^\nu)  \; p_d \cdot x_{1,6}(a,b,c,d,e,f) + (\omega_b^\mu \omega_c^\nu + \omega_c^\mu \omega_b^\nu) \; p_c \cdot x_{1,6}(b,c,d,e,f,a) \nonumber\\
& \quad + (\omega_b^\mu \omega_d^\nu + \omega_d^\mu \omega_b^\nu)  \; p_d \cdot x_{1,6}(b,c,d,e,f,a) + (\omega_c^\mu \omega_d^\nu + \omega_d^\mu \omega_c^\nu) \; p_d \cdot x_{1,6}(c,d,e,f,a,b) \,.
\end{align}

With $n_{2,6}$ in hand, we now turn our attention to $n_{1,6}$, the part of the hexagon numerator that is linear in loop momentum. It is determined by 
\begin{align}
p_a \cdot n_{1,6} (a,b,c,d,e,f) &= x_{0,6} (a,b,c,d,e,f) + p_a \cdot n_{2,6}(a,b,c,d,e,f) \cdot p_a \nn\\
&= x_{0,6} (a,b,c,d,e,f) + \frac 12  x_{1,6}(a,b,c,d,e,f) \cdot p_a\,.
\end{align}
We have met equations like this several times already; in this case, there is a slight novelty since $n_{1,6}(a,b,c,d,e,f)$ is not cyclically symmetric. Instead, the linear numerator satisfies
\begin{align}
\label{eq:n16cyclicShift}
n_{1,6}(b,c,d,e,f,a) &= n_{1,6}(a,b,c,d,e,f) - 2 p_a \cdot n_{2,6}(a,b,c,d,e,f) \nn\\ 
&= n_{1,6}(a,b,c,d,e,f) -  x_{1,6}(a,b,c,d,e,f)\,.
\end{align}
Taking this into account, the unique solution is
\begin{equation}
n_{1,6}^\mu(a,b,c,d,e,f) = \omega_a^\mu \: \eta_a +  \omega_b^\mu\: \eta_b +  \omega_c^\mu\: \eta_c +  \omega_d^\mu\: \eta_d\,,
\end{equation}
where
\begin{align}
\eta_a &= x_{0,6} (a,b,c,d,e,f) +  p_a \cdot \bigg(  \frac 12 x_{1,6}(a,b,c,d,e,f) \bigg)\,, \\
\eta_b &= x_{0,6} (b,c,d,e,f,a) +  p_b \cdot \bigg(  \frac 12 x_{1,6}(b,c,d,e,f,a) +  x_{1,6}(a,b,c,d,e,f) \bigg)\,, \\
\eta_c & = x_{0,6} (c,d,e,f,a,b) +  p_c \cdot \bigg(  \frac 12 x_{1,6}(c,d,e,f,a,b) + x_{1,6}(b,c,d,e,f,a) \nonumber \\
	& \hskip 5cm +  x_{1,6}(a,b,c,d,e,f) \bigg)\,, \\
\eta_d &= x_{0,6} (d,e,f,a,b,c) +  p_d \cdot \bigg( \frac 12 x_{1,6}(d,e,f,a,b,c) + x_{1,6}(c,d,e,f,a,b) \nonumber \\
	 & \hskip 5cm + x_{1,6}(b,c,d,e,f,a)  +  x_{1,6}(a,b,c,d,e,f) \bigg)\,.
\end{align}

We have now constructed all of the hexagon numerators other than the scalar hexagon, $n_{0,6}$. Our methods do not fix this part of the hexagon numerator at this stage, so we will continue to the heptagon. As usual, the oxidation algorithm yields a simple equation satisfied by the numerator
\begin{equation}
n_7(a,b,c,d,e,f,g;\ell) - n_7(b,c,d,e,f,g,a;\ell) = x_7(a,b,c,d,e,f,g;\ell)\,,
\end{equation}
where $x_7(a,b,c,d,e,f,g;\ell)$ is given by the familiar sum over hexagons:
\begin{align}
x_7(a,b,c,d,e,f,g;\ell) &= n_6([a,b],c,d,e,f,g;\ell) + n_6(b,[a,c],d,e,f,g;\ell) \nn\\ 
&\phantom{=} + n_6(b,c,[a,d],e,f,g;\ell) +  n_6(b,c,d,[a,e],f,g;\ell) \nn\\
&\phantom{=} +  n_6(b,c,d,e,[a,f],g;\ell) +  n_6(b,c,d,e,f,[a,g];\ell) \,.
\end{align}
Since $n_6$ contains up to two powers of loop momentum, in principle $n_7$ could contain up to three powers. However, we find that this does not occur since the terms in $x_7$ which contain two powers of loop momentum cancel out. We therefore expand 
\begin{align}
 n_7(a,b,c,d,e,f,g) &= n_{0,7}(a,b,c,d,e,f,g) + n_{1,7}(a,b,c,d,e,f,g) \cdot \ell \nn\\
& \phantom{=} + \ell\cdot n_{2,7}(a,b,c,d,e,f,g)\cdot \ell\,,
\end{align}
and find that the various different parts of the numerator satisfy
\begin{align}
2 p_a \cdot n_{2,7}(a,b,c,d,e,f,g) &= x_{1,7}(a,b,c,d,e,f,g) \,, \\
p_a \cdot n_{1,7}(a,b,c,d,e,f,g) &= x_{0,7}(a,b,c,d,e,f,g) + \frac 12 x_{1,7}(a,b,c,d,e,f,g) \cdot p_a\,.
\label{eq:walkingHeptagon}
\end{align}
These are exactly the equations the hexagon numerators satisfy. They may be solved by precisely the same methods as in \sect{SixPointSection}; although $n_{2,7}(a,b,c,d,e,f,g)$ is not permutation symmetric, it is still cyclically symmetric, which is sufficient. Meanwhile, the cyclic shift of $n_{1,7}$ is in complete analogy with \eqn{eq:n16cyclicShift}, namely
\begin{equation}
n_{1,7}(b,c,d,e,f,g,a) = n_{1,7}(a,b,c,d,e,f,g) - x_{1,7}(a,b,c,d,e,f,g)\,.
\end{equation}
At this point, we have not constructed the part of the hexagon numerator which is independent of loop momentum, so we do not have $x_{0,7}$ at our disposal. We do have $x_{1,7}$ and may therefore solve for $n_{2,7}$. It is 
\begin{align}
n_{2,7}^{\mu \nu}&(a,b,c,d,e,f,g) = \omega_a^\mu \; \omega_a^\nu \; p_a \cdot x_{1,7}(a,b,c,d,e,f,g) + \omega_b^\mu \; \omega_b^\nu \; p_b \cdot x_{1,7}(b,c,d,e,f,g,a) \nonumber \\
& \quad + \omega_c^\mu \; \omega_c^\nu \; p_c \cdot x_{1,7}(c,d,e,f,g,a,b) + \omega_d^\mu \; \omega_d^\nu \; p_d \cdot x_{1,7}(d,e,f,g,a,b,c) \nonumber \\
& \quad + (\omega_a^\mu \omega_b^\nu + \omega_b^\mu \omega_a^\nu)  \; p_b \cdot x_{1,7}(a,b,c,d,e,f,g) + (\omega_a^\mu \omega_c^\nu + \omega_c^\mu \omega_a^\nu) \; p_c \cdot x_{1,7}(a,b,c,d,e,f,g) \nonumber\\
& \quad + (\omega_a^\mu \omega_d^\nu + \omega_d^\mu \omega_a^\nu)  \; p_d \cdot x_{1,7}(a,b,c,d,e,f,g) + (\omega_b^\mu \omega_c^\nu + \omega_c^\mu \omega_b^\nu) \; p_c \cdot x_{1,7}(b,c,d,e,f,g,a) \nonumber\\
& \quad + (\omega_b^\mu \omega_d^\nu + \omega_d^\mu \omega_b^\nu)  \; p_d \cdot x_{1,7}(b,c,d,e,f,g,a) + (\omega_c^\mu \omega_d^\nu + \omega_d^\mu \omega_c^\nu) \; p_d \cdot x_{1,7}(c,d,e,f,g,a,b) \,.
\end{align}
A non-trivial property satisfied by this numerator is that upon integrand reduction to pentagons, it cancels against $n_{2,6}$; this property is analogous to what we saw at six points, in Eq.~\eqref{eq:n16sol2}. In fact, if we impose that on integrand reduction to pentagons, the terms linear in $\ell$ also cancel, we can determine $n_{1,7}$. It is given by
\begin{multline}
n_{1,7}(a,b,c,d,e,f,g) = \frac{-1}{r(ab|cd|e|f|g)} \left(\frac{n_{1,6}([a,b],c,d,e,f,g)}{s_{ab}} r(a{+}b|cd|e|f|g)  \right. \\
\left. + \frac{n_{1,6}(a,b,[c,d],e,f,g)}{s_{cd}} r(ab|c{+}d|e|f|g))  + \frac{n_{1,5}([a,b],[c,d],e,f,g)}{s_{ab}s_{cd}}\right)\,.
\end{multline}
It is non-trivial that this expression for $n_{1,7}(a,b,c,d,e,f,g)$ has the property that all the Jacobi identities are satisfied. 

Encouraged by this success, we recall that \eqn{eq:walkingHeptagon} shows that $n_{1,7}$ is determined by the scalar parts of the hexagons; turning this around, the scalar hexagons are constrained by knowledge of $n_{1,7}$. Let us begin by understanding how much freedom there is at this point in the scalar hexagon numerators. Since we know all the scalar pentagons, we can use Jacobi moves to bring a given scalar hexagon to a canonical order. We choose to place the unique massive leg of the hexagon in the first entry of the function; we may order the rest of the legs numerically. There are 21 choices of the massive leg, so as it stands there are 21 functions of the form $n_{0,6}([a,b],c,d,e,f,g)$ to be determined; all other scalar hexagons can be determined from these 21 functions and the scalar pentagons. The structure of the equations relating the scalar hexagon numerators $n_{0,6}$ to the linear heptagon numerator $n_{1,7}$ is as follows. Define
\begin{align}
s_1(a,b,c,d,e,f,g) &= p_b \cdot n_{1,6}([a,c],d,e,f,g,b) + p_{bc} \cdot n_{1,6}([a,d],e,f,g,b,c) \nn\\ 
& + p_{bcd} \cdot n_{1,6}([a,e],f,g,b,c,d) + p_{bcde} \cdot n_{1,6}([a,f],g,b,c,d,e) \nn\\
&+ p_{bcdef} \cdot n_{1,6}([a,g],b,c,d,e,f) 
\end{align}
and
\begin{align}
s_2(a,b,c,d,e,f,g) &= p_b \cdot n_{2,6}([a,c],d,e,f,g,b) \cdot p_b + p_{bc} \cdot n_{2,6}([a,d],e,f,g,b,c) \cdot p_{bc} \nn\\
& + p_{bcd} \cdot n_{2,6}([a,e],f,g,b,c,d) \cdot p_{bcd} + p_{bcde} \cdot n_{2,6}([a,f],g,b,c,d,e) \cdot p_{bcde} \nn\\
& + p_{bcdef} \cdot n_{2,6}([a,g],b,c,d,e,f) \cdot  p_{bcdef}  \,.
\end{align}
These quantities simply account for cycling the hexagon numerators to a canonical order. We then find
\begin{align}
&n_{0,6}([a,b],c, d,e,f,g) + n_{0,6}([a,c],d,e,f,g,b) + n_{0,6}([a,d],e,f,g,b,c) \nn\\
& + n_{0,6}([a,e],f, g,b,c,d) + n_{0,6}([a,f],g,b,c,d,e)+n_{0,6}([a,g],b,c,d,e,f)  \nn\\
&\hskip 2.5cm = p_a \cdot n_{1,7}(a,b,c,d,e,f,g) - \frac 12 p_a \cdot x_{1,7}(a,b,c,d,e,f,g) -s_1(a,b,c,d,e,f,g)\nn\\
&\hskip 2.5cm \phantom{=}- s_2(a,b,c,d,e,f,g)\,.
\end{align}
Note that the unknowns appearing in this equation include all possible choices of scalar hexagon numerator whose massive leg includes particle $a$. There can therefore be at most seven independent equations, corresponding to the different possible choices for leg $a$. In fact, only six of these equations are independent. We may therefore use these equations to solve for the six different numerators whose massive leg include a given particle, for example, particle 7. Thus,
\begin{align}
& n_{0,6}([a,7],b,c,d,e,f)  \nonumber \\
&\hskip 2cm =p_a \cdot n_{1,7}(a,b,c,d,e,f,7) - \frac 12 p_a \cdot x_{1,7}(a,b,c,d,e,f,7)  - s_1(a,b,c,d,e,f,7)\nonumber \\ 
&\hskip 2cm  \phantom{=}- s_2(a,b,c,d,e,f,7) - n_{0,6}([a,b],c,d,e,f,7) - n_{0,6}([a,c],d,e,f,7,b) \nonumber \\ 
&\hskip 2cm \phantom{=}- n_{0,6}([a,d],e,f,7,b,c) - n_{0,6}([a,e],f,7,b,c,d) - n_{0,6}([a,f],7,b,c,d,e)\,.
\end{align}
There are 15 scalar hexagon numerators still to be fixed. We will fix them by integrand reduction. But before we do so, we must compute the family of scalar heptagon numerators which correspond to the space of hexagon numerators we are considering.

Given a scalar hexagon numerator, it is straightforward to compute the scalar heptagon numerator. We use the reflection symmetry of the heptagon to do so; one convenient method is to note that
\begin{equation}
n_7(a,b,c,d,e,f,g;\ell) - n_7(d,c,b,a,g,f,e;\ell) = y_7(a,b,c,d,e,f,g;\ell)\,,
\end{equation}
where $y_7$ is the following sum over hexagon numerators
\begin{align}
&y_7(a,b,c,d,e,f,g;\ell) = n_6([a,b],c,d,e,f,g;\ell) + n_6(b,[a,c],d,e,f,g;\ell) \nonumber \\ 
&\hskip 2cm+ n_6(b,c,[a,d],e,f,g;\ell) + n_6([b,c],d,a,e,f,g;\ell) + n_6(c,[b,d],a,e,f,g;\ell) \nonumber \\ 
&\hskip 2cm+ n_6([c,d],b,a,e,f,g;\ell) + n_6(d,c,b,a,[e,f],g;\ell) + n_6(d,c,b,a,f,[e,g];\ell) \nonumber \\ 
&\hskip 2cm + n_6(d,c,b,a,[f,g],e;\ell)\,.
\end{align}
The symmetries of the heptagon require
\begin{equation}
n_7(d,c,b,a,g,f,e;\ell) = - n_7(a,b,c,d,e,f,g;-\ell+p_e + p_f + p_g)\,,
\end{equation}
so that we can deduce a very simple expression for the scalar heptagon numerator $n_{0,7}$ in terms of the scalar hexagons and other known quantities:
\begin{align}
n_{0,7}&(a,b,c,d,e,f,g)  \nonumber \\
& = \frac 12 \big( y_{0,7}(a,b,c,d,e,f,g) - p_{efg} \cdot n_{1,7}(a,b,c,d,e,f,g) - p_{efg} \cdot n_{2,7}(a,b,c,d,e,f,g)\cdot p_{efg} \big)\,.
\end{align}

So we have constructed a 15-parameter family of numerators for the seven-point amplitude that satisfies all the Jacobi relations. We now want to exploit the freedom in our construction to make contact with the known expressions for the MHV boxes on integrand reduction. One simple property of the box coefficients is that the three-mass-hard boxes all vanish \cite{Bern:1994zx,Bern:1998sv}. We find that this requirement leads to 13 independent equations of the form
\begin{align}
0 &= n_{0,7}(a,b,c,d,e,f,g)\: r(ab|cd|ef|g) + \frac{n_{0,6}([a,b],c,d,e,f,g)}{s_{ab}} r(a{+}b|cd|ef|g) \nn\\
&\phantom{=} + \frac{n_{0,6}(a,b,[c,d],e,f,g)}{s_{cd}} r(ab|c{+}d|ef|g) 
+ \frac{n_{0,6}(a,b,c,d,[e,f],g)}{s_{ef}} r(ab|cd|e{+}f|g) \nn\\
&\phantom{=} + \frac{n_{0,5}([a,b],[c,d],e,f,g)}{s_{ab}s_{cd}} r(a{+}b|c{+}d|ef|g) 
+ \frac{n_{0,5}([a,b],c,d,[e,f],g)}{s_{ab}s_{ef}} r(a{+}b|cd|e{+}f|g) \nn\\
&\phantom{=} + \frac{n_{0,5}(a,b,[c,d],[e,f],g)}{s_{cd}s_{ef}} r(ab|c{+}d|e{+}f|g) 
+ \frac{n_{4}([a,b],[c,d],[e,f],g)}{s_{ab}s_{cd}s_{ef}} \,.
\end{align}
This is a linear system of equations which may be solved to fix 13 of the remaining unknown scalar hexagon numerators. Two numerators remain free. Indeed, we have more physical information at our disposal: we know that the two-mass-hard boxes vanish, while the two-mass-easy and one-mass boxes take on particular values. However, all of these requirements are automatically satisfied. We are therefore left with two functional parameters which may be chosen arbitrarily. This completes our construction of the seven-point amplitude.


\section{Beyond MHV Amplitudes}
\label{NMHVSection}

In the previous sections of this paper, we have exploited information originating in the self-dual sector of Yang-Mills theory as the basis for building complete sets of colour-dual numerators. Since the dimension-shifting formula is limited (as far as we know) to the MHV amplitudes, we present here a more general construction.  We determine a linear map that fixes the scalar part $n_0$ of the colour-dual numerators from unitarity cuts  \cite{Bern:1994zx,Bern:1998sv}, subject to the assumption that all $\ell$ dependence in our numerators cancels upon reduction to pentagons, as happened in the MHV case. The fact that this linear map has solutions is due to nontrivial identities between the unitarity cuts. The $\ell$ dependence is then fixed by the scalar part using the integrand oxidation methods described in previous sections. In the five-point case, the linear map is simple enough to be solved analytically; for six and seven points, we outline its construction and find solutions. We see no obstruction to the extension of this method to higher multiplicities.

\subsection*{Five Points}

At five points, there are only one-mass boxes available. In the BDDK presentation of the amplitude (i.e., after integrand reduction to boxes, throwing away the parity-odd terms) the box coefficients are given by
\begin{equation}
\label{eq:5ptMassiveBox}
 b(12|3|4|5) = \frac 14 \delta^8(Q) \frac{\tr ((1{+}2) 345)}{\ab{12} \ab{23} \ab {34} \ab {45} \ab{51} }\,.
\end{equation}
Our goal is to construct a set of colour-dual numerators which integrand-reduce to these one-mass boxes. In other words, we wish to express the integrand of the colour-ordered amplitude as
\begin{equation}
\mathcal{I} = i\: n_5(1,2,3,4,5)
\begin{minipage}[c]{0.2\linewidth}
\centering
\includegraphics[scale=0.3]{pentagon}
\end{minipage}
+ i\sum_{\mathbb{Z}_5}  n_4([a,b],c,d,e)
\begin{minipage}[c]{0.2\linewidth}
\centering
\includegraphics[scale=0.3]{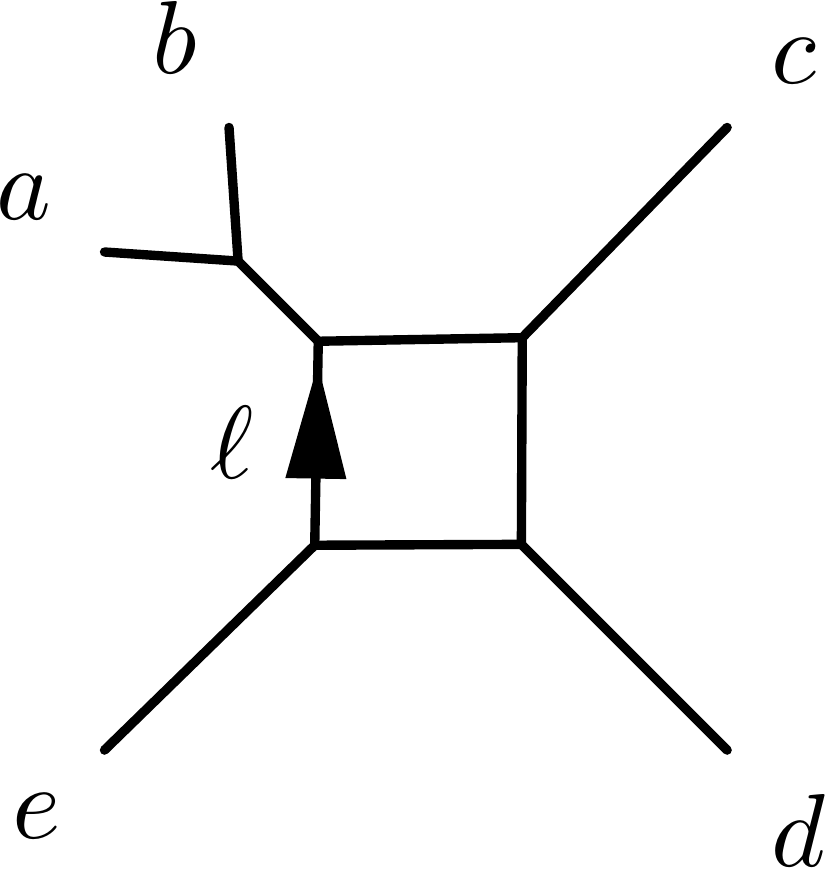}
\end{minipage}, 
\end{equation}
where the sum is over the five cyclic permutations of $(1,2,3,4,5)$. We will refer to this presentation of the integrand as the colour-dual presentation. Notice that we have assumed that the pentagon numerator is independent of loop momentum; this is consistent with our general expectations but can be thought of as an Ansatz for the purpose of the present discussion.

Now, the colour-dual presentation must equal the BDDK presentation on integrand reduction, which leads to 
\begin{equation}
\label{eq:5ptReduction}
-i\:b(12|3|4|5) = n_5(1,2,3,4,5)\: r(12|3|4|5) + \frac{n_5(1,2,3,4,5) - n_5(2,1,3,4,5)}{s_{12}}\,.
\end{equation}
This equation is not especially transparent, but we may derive a simpler equation as follows: switch legs $1$ and $2$ in Eq.~\eqref{eq:5ptReduction} to find
\begin{equation}
\label{eq:5ptReductionSwap}
-i \:b(21|3|4|5) = n_5(2,1,3,4,5)\: r(21|3|4|5) + \frac{n_5(2,1,3,4,5) - n_5(1,2,3,4,5)}{s_{21}}\,,
\end{equation}
and add this to Eq.~\eqref{eq:5ptReduction}. The result is obviously
\begin{align}
\label{eq:5ptMaster}
n_5(1,2,3,4,5)&\: r(12|3|4|5) + n_5(2,1,3,4,5)\: r(21|3|4|5) \nonumber \\
&= -\frac i4 \delta^8(Q)\, \tr ((1{+}2) 345) \left( \frac{1}{\ab{12} \ab{23} \ab {34} \ab {45} \ab{51} } + \frac{1}{\ab{21} \ab{13} \ab {34} \ab {45} \ab{52} }\right) \nonumber \\ 
&\equiv X(\{1,2\},3,4,5).
\end{align}
This equation is a kind of analogue of the Jacobi equation; the two terms on the left differ by an adjacent transposition. We may therefore consider moving one leg, for example leg $1$, around the loop. In this way, we will obtain a simple equation for the numerator $n_5(1,2,3,4,5)$. It is straightforward to handle the appearance of the reduction coefficient factors. We find
\begin{align}
&n_5(1,2,3,4,5) \left( r(12|3|4|5) - r(51|2|3|4) \frac{r(21|3|4|5)}{r(13|4|5|2)}\frac{r(31|4|5|2)}{r(14|5|2|3)} \frac{r(41|5|2|3)}{r(15|2|3|4)} \right) \nonumber \\
&\; = X(\{1,2\},3,4,5) - X(\{1,3\},4,5,2) \frac{r(21|3|4|5)}{r(13|4|5|2)}+  X(\{1,4\},5,2,3) \frac{r(21|3|4|5)}{r(13|4|5|2)} \frac{r(31|4|5|2)}{r(14|5|2|3)}  \nonumber \\
&\quad-X(\{1,5\},2,3,4) \frac{r(21|3|4|5)}{r(13|4|5|2)} \frac{r(31|4|5|2)}{r(14|5|2|3)}\frac{r(41|5|2|3)}{r(15|2|3|4)} .
\end{align}
We have checked that the solution of this equation is the familiar Carrasco-Johansson numerator
\begin{equation}
n_5(1,2,3,4,5) = -i \delta^8(Q) \frac{[12][23][34][45][51]}{\trf{1234}}.
\end{equation}
Thus, we have recovered the known colour-dual set of numerators without starting from any particularly nice set of boxes.

\subsection*{Six Points}

This method generalises to higher-point amplitudes. At six points, we can again match the colour-dual presentation of the amplitude to all of the box cuts:
\begin{align}
\label{eq:hex2mh}
\text{2-mass hard:    }&	-i\: b(12|34|5|6) = n_{0,6}(1,2,3,4,5,6)\: r(12|34|5|6) \nonumber \\
	& \hskip .5cm  + \frac{ n_{0,5}([1,2],3,4,5,6)}{s_{12}}  r(1{+}2|34|5|6) + \frac{n_{0,5}(1,2,[3,4],5,6)}{s_{34}}  r(12|3{+}4|5|6) \nonumber \\
	& \hskip .5cm + \frac{ n_{0,4}([1,2],[3,4],5,6)}{s_{12}s_{34}} \,,\\
\label{eq:hex2me}
\text{2-mass easy:    }& 	-i\: b(12|3|45|6) = n_{0,6}(1,2,3,4,5,6) \: r(12|3|45|6) \nonumber \\
	& \hskip .5cm + \frac{ n_{0,5}([1,2],3,4,5,6)}{s_{12}} r(1{+}2|3|45|6) + \frac{ n_{0,5}(1,2,3,[4,5],6)}{s_{45}}  r(12|3|4{+}5|6) \nonumber \\
	& \hskip .5cm + \frac{ n_{0,4}([1,2],3,[4,5],6)}{s_{12}s_{45}} \,, \\
\label{eq:hex1m}
\text{1-mass:    } &		-i\: b(123|4|5|6) = n_{0,6}(1,2,3,4,5,6) \: r(123|4|5|6) \nonumber\\ 
	& \hskip .5cm + \frac{ n_{0,5}([1,2],3,4,5,6}{s_{12}}  r((1{+}2)3|4|5|6) + \frac{ n_{0,5}(1,[2,3],4,5,6)}{s_{23}}  r(1(2{+}3)|4|5|6) \nonumber \\
	& \hskip .5cm + \frac{ n_{0,4}([1,[2,3]],4,5,6)}{s_{123}s_{23}} + \frac{n_{0,4}([[1,2],3],4,5,6)}{s_{123}s_{12}} \,.
\end{align}
In these equations, the box coefficients $b(\ldots)$ are known explicitly, regardless of whether we are talking about MHV or NMHV amplitudes.  Note that, motivated by our experience with the MHV six- and seven-point amplitudes, we have made the crucial assumption that the loop momentum dependence of the colour-dual numerators drops out after reduction to pentagons. Therefore, only the scalar numerators $n_0$ remain, and the final reduction from pentagons to boxes is simply stated using the reduction coefficient $r(\ldots)$. We also impose reflection symmetry and the vanishing of all triangle numerators, 
\begin{align}
\label{eq:hexreflection}
	n_{0,6}(a,b,c,d,e,f) & = n_{0,6}(f,e,d,c,b,a)\,, \\
\label{eq:hextri}
	n_{0,3}(a,b,c) & = 0\,.
\end{align}
Using the Jacobi identities, we can rewrite $n_{0,5}$,  $n_{0,4}$, and $n_{0,3}$ in terms of $n_{0,6}$. Eqs.~\eqref{eq:hex2mh}--\eqref{eq:hextri} for all permutations of external legs then provide a linear system constraining the 720 unknown quantities $n_{0,6}(a,b,c,d,e,f)$. This system has rank 719, subject to a number of consistency conditions on the box coefficients $b(\ldots)$ that are too lengthy to give here. We have verified that these consistency conditions are indeed satisfied by the box coefficients for both MHV and NMHV amplitudes, and we have solved the system for particular (generic) choices of kinematics.

With $n_{0,6}$ in hand, it is now straightforward to reconstruct all of the $\ell$ dependence of the colour-dual numerators using the integrand oxidation procedure discussed in previous sections. In the MHV case, the one remaining degree of freedom is equivalent to Yuan's $\kappa$ parameter \cite{Yuan:2012rg}, and leads to a quadratic term in the hexagon numerator equal to
\begin{equation}
	n_{2,6}^{\mu\nu}(1,2,3,4,5,6) = -3i\kappa\: \delta^8(Q) \:\eta^{\mu\nu}\,.
\end{equation}
This is inconsistent with our requirement that the loop momentum dependence drops out after integrand reduction to pentagons, and indeed the reconstructed numerators only give the correct amplitude for the choice $\kappa = 0$. The vanishing of $n_{2,6}$ also eliminates the remaining degree of freedom in the NMHV amplitude.

\subsection*{Seven Points}

The same setup works again at seven points. We start by writing down all of the box cut constraints, which relate the 5040 unknowns $n_{0,7}(a,b,c,d,e,f,g)$ to known box coefficients. Together with reflection symmetry and vanishing of triangle numerators, this linear system has rank 5019, again subject to consistency conditions on the box coefficients $b(\ldots)$. We have verified that the box coefficients in the MHV and NMHV amplitudes satisfy the conditions, and we have explicitly solved the system for 5019 of the $n_{0,7}(a,b,c,d,e,f,g)$ in terms of the 21 others. 

The $\ell$ dependence of the numerators can be constructed using the integrand oxidation procedure. As at six points, the story is not finished; the condition that $\ell$ drops out of the numerators after reduction to pentagons provides a further 19 constraints, and we are left with a 2-parameter family of colour-dual numerators for seven-point MHV and NMHV amplitudes. For the MHV amplitude, this matches exactly the freedom found by the analysis in \sect{SevenPointSection}.

We have thus established definitively the existence of colour-dual numerators for all one-loop six- and seven-point amplitudes in $\mN=4$ SYM. Under our assumptions, the box numerators are unique and, for the MHV amplitude, match the box numerators obtained from the self-dual amplitudes.


\section{Gravity Amplitudes}
\label{Gravity}

One of the most important properties of the BCJ numerators that we constructed in the previous sections is the cancellation of the loop momentum dependence upon integrand reduction to pentagons. In this section, we will see that the same happens when we ``square" the $\mN=4$ SYM numerators to obtain the corresponding one-loop amplitudes in $\mN=8$ supergravity, following the BCJ double copy prescription, Eq.~\eqref{eq:loopdoublecopy}. 

\subsection*{Six Points}

The six-point gauge theory amplitude was constructed in Section~\ref{SixPointSection} in the MHV sector and Section~\ref{NMHVSection} in the NMHV sector. It is the first case where our BCJ numerators depend on the loop momentum. The dependence is linear,
\begin{equation}
n_6(1,2,3,4,5,6;\ell) = n_{0,6}(1,2,3,4,5,6) + n_{1,6}(1,2,3,4,5,6) \cdot \ell\,.
\end{equation}
The dependence on $\ell$ is such that, upon integrand reduction to a pentagon, it cancels against that of the pentagon numerator. For instance, take the coefficient of the pentagon integral $I_5^{(1+2)|3|4|5|6}$ in the partial amplitude; the piece proportional to $\ell$ is
\begin{equation}
n_{1,6}^\mu(1,2,3,4,5,6) \:r(12|3|4|5|6) + \frac{n_{1,5}^\mu([1,2],3,4,5,6)}{s_{12}} =0\,.
\label{eq:6ptcancelYM}
\end{equation}
This cancellation can be traced back to the system of Jacobi identities that determines $n_{1,5}$ and $n_{1,6}$. For a simple illustration of how tightly connected integrand reduction and Jacobi identities are, consider the following consistency check. Using Eq.~\eqref{eq:6ptcancelYM}, we get
\begin{align}
n_{1,6}^\mu(1,2,3,4,5,6) &- n_{1,6}^\mu(2,1,3,4,5,6)  \nonumber \\
&= -  \frac{n_{1,5}^\mu([1,2],3,4,5,6)}{s_{12}} \left( \frac{1}{r(12|3|4|5|6)} + \frac{1}{r(21|3|4|5|6)} \right)  \nonumber \\
&= n_{1,5}^\mu([1,2],3,4,5,6)\,,
\end{align}
where we imposed $n_{1,5}^\mu([2,1],3,4,5,6)=-n_{1,5}^\mu([1,2],3,4,5,6)$ and used the reduction coefficients given in \eqn{hexreduction}.

Let us now consider the cancellation in the gravity amplitude obtained through the double copy, focusing again on the coefficient of the integral $I_5^{(1+2)|3|4|5|6}$. The numerator $n_6$ is linear in loop momentum, so its square will have a quadratic piece, a linear piece and a scalar piece. The contribution of the quadratic piece is
\begin{align}
&n_{1,6}^\mu(1,2,3,4,5,6)\: n_{1,6}^\nu(1,2,3,4,5,6)\: r(12|3|4|5|6) \nonumber \\
&+ n_{1,6}^\mu(2,1,3,4,5,6)\: n_{1,6}^\nu(2,1,3,4,5,6)\: r(21|3|4|5|6)  \nonumber \\
& \qquad = \frac{n_{1,5}^\mu([1,2],3,4,5,6)\: n_{1,5}^\nu([1,2],3,4,5,6)}{s_{12}^2} \left( \frac{1}{r(12|3|4|5|6)} + \frac{1}{r(21|3|4|5|6)} \right)  \nonumber \\
& \qquad = -\frac{n_{1,5}^\mu([1,2],3,4,5,6)\: n_{1,5}^\nu([1,2],3,4,5,6)}{s_{12}} \,,
\end{align}
where we used the cancellation in gauge theory, Eq.~\eqref{eq:6ptcancelYM}. Notice that, because there is no colour ordering in gravity, we had to consider the two terms $(12|3|4|5|6)$ and $(21|3|4|5|6)$. Thus the hexagon $\ell^\mu\ell^\nu$ contribution cancels the pentagon contribution. Likewise, we find a cancellation for the piece linear in $\ell$,
\begin{align}
&n_{1,6}^\mu(1,2,3,4,5,6)\: n_{0,6}(1,2,3,4,5,6)\:  r(12|3|4|5|6) \nonumber \\
&+ n_{1,6}^\mu(2,1,3,4,5,6)\: n_{0,6}(2,1,3,4,5,6)\:  r(21|3|4|5|6)  \nonumber \\
& \qquad = -\frac{n_{1,5}^\mu([1,2],3,4,5,6)}{s_{12}} \big( n_{0,6}(1,2,3,4,5,6)-n_{0,6}(2,1,3,4,5,6) \big)  \nonumber \\
& \qquad = -\frac{n_{1,5}^\mu([1,2],3,4,5,6)\: n_{0,5}([1,2],3,4,5,6)}{s_{12}} \,.
\end{align}

We conclude that the cancellation of $\ell$ dependence in gravity follows from the analogous cancellation in gauge theory. Let us emphasise that this does not mean we can just drop the linear terms $n_{1,5}\cdot \ell$ and $n_{1,6}\cdot \ell$ of the numerators in all the cubic diagrams. Those terms ensure that the integrand reduction has the correct global properties. Consider, for instance, the integrand reduction in a gauge theory partial amplitude. Placing the loop momentum between legs 6 and 1, the hexagon integrand reduces to six distinct pentagon integrands, as in the expression \eqref{eq:sumrP}. For five of those pentagons, the reduction proceeds as in the case \eqref{eq:6ptcancelYM}, but for the pentagon $I_5^{(6+1)|2|3|4|5}$, we have
\begin{align}
&\hskip -2cm n_6(1,2,3,4,5,6;\ell) \: r(61|2|3|4|5) + \frac{n_5([6,1],2,3,4,5;\ell-p_6)}{s_{61}} \nonumber \\
&=  n_{0,6}(6,1,2,3,4,5)\:r(61|2|3|4|5) + \frac{n_{0,5}([6,1],2,3,4,5)}{s_{61}} \,,
\end{align}
where we recall that $n_{6}$ is cyclically symmetric, so that
\begin{align}
n_{0,6}(1,2,3,4,5,6) =  n_{0,6}(6,1,2,3,4,5) - n_{1,6}(1,2,3,4,5,6) \cdot p_6\,
\end{align}
and
\begin{align}
n_{1,6}^\mu(1,2,3,4,5,6) \,r(61|2|3|4|5) + \frac{n_{1,5}^\mu([6,1],2,3,4,5) }{s_{61}} = 0 \,.
\end{align}
So we see that the naive substitution $n_6(1,2,3,4,5,6;\ell) \to n_{0,6}(1,2,3,4,5,6)$ does not correctly reproduce all six pentagons simultaneously.

\subsection*{Seven Points}

The situation is entirely analogous at seven points. Now the `master numerator'  $n_7$, constructed in Section~\ref{SevenPointSection} (\ref{NMHVSection}) in the MHV (NMHV) sector, is a quadratic polynomial in the loop momentum,
\begin{align}
n_7(1,2,3,4,5,6,7;\ell) &=  n_{0,7}(1,2,3,4,5,6,7) + n_{1,7}(1,2,3,4,5,6,7) \cdot \ell \nonumber \\
&\phantom{=}+ \ell \cdot n_{2,7}(1,2,3,4,5,6,7) \cdot \ell \,.
\end{align}
We have to consider two types of pentagons: one-mass and two-mass.

Starting with the two-mass pentagons, and focusing on the coefficient of the integral $I_5^{(1+2)|(3+4)|5|6|7}$, the cancellation of loop momentum dependence upon integrand reduction in gauge theory is expressed as:
\begin{align}
& n_{2,7}^{\mu\nu}(1,2,3,4,5,6,7)\: r(12|34|5|6|7) + \frac{n_{2,6}^{\mu\nu}([1,2],3,4,5,6,7)}{s_{12}} r(1{+}2|34|5|6|7) \nonumber \\
&+ \frac{n_{2,6}^{\mu\nu}(1,2,[3,4],5,6,7)}{s_{34}} r(12|3{+}4|5|6|7) =0\,,
\label{eq:7ptcancelYM2mass2}
\end{align}
and
\begin{align}
& n_{1,7}^\mu(1,2,3,4,5,6,7)\: r(12|34|5|6|7) + \frac{n_{1,6}^\mu([1,2],3,4,5,6,7)}{s_{12}} r(1{+}2|34|5|6|7) \nonumber \\
&+ \frac{n_{1,6}^\mu(1,2,[3,4],5,6,7)}{s_{34}} r(12|3{+}4|5|6|7) + \frac{n_{1,5}^\mu([1,2],[3,4],5,6,7)}{s_{12}s_{34}} =0 \,,
\label{eq:7ptcancelYM2mass1}
\end{align}
where we considered both the quadratic and the linear pieces, respectively; recall that $n_{2,5}=0$. Since $n_7$ is a quadratic polynomial in $\ell$, its square is a quartic polynomial. The cancellation in gravity corresponds to the fact that the $\ell$ dependence drops out in the following expression:
\begin{align}
& n_{7}^2(1,2,3,4,5,6,7;\ell)\: r(12|34|5|6|7) + n_{7}^2(2,1,3,4,5,6,7;\ell)\: r(21|34|5|6|7) \nonumber \\
&+ n_{7}^2(1,2,4,3,5,6,7;\ell)\: r(12|43|5|6|7) + n_{7}^2(2,1,4,3,5,6,7;\ell)\: r(21|43|5|6|7) \nonumber \\
&+ \frac{n_6^2([1,2],3,4,5,6,7;\ell)}{s_{12}} r(1{+}2|34|5|6|7) + \frac{n_6^2([1,2],4,3,5,6,7;\ell)}{s_{12}} r(1{+}2|43|5|6|7) \nonumber \\
&+ \frac{n_6^2(1,2,[3,4],5,6,7;\ell)}{s_{34}} r(12|3{+}4|5|6|7) + \frac{n_6^2(2,1,[3,4],5,6,7;\ell)}{s_{34}} r(21|3{+}4|5|6|7) \nonumber \\
&+ \frac{n_5^2([1,2],[3,4],5,6,7;\ell)}{s_{12}s_{34}} \,.
\label{eq:7ptcancelgrav2mass}
\end{align}
Notice that this requires the cancellation of pieces which are quartic, cubic, quadratic and linear in $\ell$. This all follows from \eqref{eq:7ptcancelYM2mass2} and \eqref{eq:7ptcancelYM2mass1}, together with the Jacobi identities.

For the other type of pentagon, namely the one-mass pentagon, let us consider the coefficient of the integral $I_5^{(1+2+3)|4|5|6|7}$ in gauge theory. The cancellations are
\begin{align}
& n_{2,7}^{\mu\nu}(1,2,3,4,5,6,7)\: r(123|4|5|6|7) + \frac{n_{2,6}^{\mu\nu}([1,2],3,4,5,6,7)}{s_{12}}  r((1{+}2)3|4|5|6|7) \nonumber \\
&+ \frac{n_{2,6}^{\mu\nu}(1,[2,3],4,5,6,7)}{s_{34}} r(1(2{+}3)|4|5|6|7) =0\,,
\label{eq:7ptcancelYM1mass2}
\end{align}
and
\begin{align}
& n_{1,7}^{\mu}(1,2,3,4,5,6,7)\: r(123|4|5|6|7)  \nonumber \\
&+ \frac{n_{1,6}^{\mu}([1,2],3,4,5,6,7)}{s_{12}}  r((1{+}2)3|4|5|6|7) + \frac{n_{1,6}^{\mu}(1,[2,3],4,5,6,7)}{s_{34}} r(1(2{+}3)|4|5|6|7)  \nonumber \\
& + \frac{n_{1,5}^\mu([[1,2],3],4,5,6,7)}{s_{12}s_{123}}+ \frac{n_{1,5}^\mu([1,[2,3]],4,5,6,7)}{s_{23}s_{123}} =0\,.
\label{eq:7ptcancelYM1mass1}
\end{align}
For the gravity amplitude, these relations imply that the $\ell$ dependence drops out in the following quantity:
\begin{align}
& n_{7}^2(1,2,3,4,5,6,7;\ell)\: r(123|4|5|6|7) + n_{7}^2(2,1,3,4,5,6,7)\: r(213|4|5|6|7) \nonumber \\
&+ \frac{n_6^2([1,2],3,4,5,6,7;\ell)}{s_{12}}  r((1{+}2)3|4|5|6|7) + \frac{n_6^2(3,[1,2],4,5,6,7;\ell)}{s_{12}} r(3(1{+}2)|4|5|6|7) \nonumber \\
&+ \frac{n_5^2([[1,2],3],4,5,6,7;\ell)}{s_{12}s_{123}} + \textrm{cyclic}(123) \,,
\label{eq:7ptcancelgrav1mass}
\end{align}
where one should add two copies of the whole expression corresponding to the other cyclic permutations of legs 1, 2 and 3. All the terms contribute to the coefficient of the pentagon integral $I_5^{(1+2+3)|4|5|6|7}$.


\section{Conclusion}
\label{ConclusionSection}

We have developed a systematic method of determining BCJ numerators for one-loop amplitudes. The strength of this method is that it makes use of the global constraints on the loop momentum dependence of the numerators imposed by the Jacobi identities. In fact, one of the motivations for this work was the discussion by Yuan \cite{Yuan:2012rg} of how a global constraint prevented the construction of six-point BCJ numerators independent of loop momentum. The global constraints allow us to build higher-polygon numerators starting with the box numerators, whence our terminology of ``integrand oxidation," as opposed to reduction.

The specific input from $\mN=4$ SYM theory that was useful is that it is possible to consider vanishing bubble and triangle numerators, and box numerators independent of the loop momentum. While we do not present a proof, our results give strong evidence that colour-kinematics duality holds for one-loop $\mN=4$ SYM amplitudes at any multiplicity and in any R-sector.

We proposed a simple prescription to compute box numerators in the MHV case, using a very interesting connection between MHV amplitudes and self-dual gauge theory proposed in \cite{Bern:1996ja}. We believe that this is a canonical choice of box numerators with nice properties. Recall that, beyond four points, it automatically prevented the `master numerator' $n_m$ from depending on the highest possible power of the loop momentum, which would have been $m-4$. The connection to the self-dual sector is suggestive, but we should also note the relation to the so-called MHV polygons proposed in \cite{ArkaniHamed:2010gh}. The same set of six-point box numerators can be obtained by considering a simple form of MHV polygons which makes the cyclic symmetry manifest \cite{Yuan:2012rg}. Beyond MHV amplitudes, we found that the numerators of box diagrams in a colour-dual expression for NMHV amplitudes were unique. This may be understood via the uniqueness of the counterterm Lagrangian which absorbs the one-loop UV divergences of super-Yang-Mills theory in eight dimensions, as pointed out by Schabinger~\cite{Schabinger:2011kb} following work of Stieberger and Taylor~\cite{Stieberger:2006bh}.\footnote{We thank R. Schabinger for enlightening correspondence on this point.} It is likely that these boxes, and possibly other aspects of the Jacobi identities are also closely related to the structure of one-loop superstring amplitudes studied in~\cite{Mafra:2012kh}.

In addition, we described a procedure to construct colour-dual numerators from knowledge of unitarity cuts alone. Using this procedure, we obtained colour-dual forms for six- and seven-point MHV and NMHV amplitudes in $\mN=4$ SYM. In principle, we believe this will work for any multiplicity and in any R sector. More work is needed to fully understand the structure of the equations that must be solved; in particular, there is a set of constraints that must be satisfied by the unitarity cuts of an amplitude for a colour-dual form to exist. It would also be interesting to determine simple analytic forms for the numerators.

A very clear property of the loop momentum dependence of the numerators is that it cancels out upon integrand reduction to pentagons. This cancellation shows that the Jacobi identities are tightly intertwined with the integrand reduction coefficients. These in turn lead to cancellations in one-loop $\mN=8$ supergravity amplitudes. While the ultraviolet properties of one-loop $\mN=8$ amplitudes have long been understood by other methods, here we have seen explicitly their intricate relationship with the colour-kinematics duality. It is of obvious interest to investigate the extension of this relationship to higher-loop amplitudes.

As a side result of our investigations, we obtained simple trace-based formulas for integral reduction to pentagons and to boxes, which are more convenient in applications using the spinor-helicity formalism than the equivalent determinant-based formulas \cite{Melrose:1965kb,Bern:1992em}. We hope that these results may find a more general use.

To conclude, our goal was to make explicit the linear map that puts a one-loop amplitude into a colour-kinematics dual form. We expect that the main idea explored here of using the global constraints on loop momentum dependence will find use beyond the one-loop level.


\appendix

\acknowledgments

It is a pleasure to thank Simon Badger, Zvi Bern, Rutger Boels, Simon Caron-Huot, Poul Henrik Damgaard, Reinke Isermann, Robert Schabinger, Oliver Schlotterer, Pierre Vanhove and Yang Zhang for useful discussions. We are especially grateful to Henrik Johansson for emphasising the relevance of the dimension-shifting formula to us.

\end{document}